\documentclass[amsmath,amssymb,showkeys,superscriptaddress,floatfix,aps,prd,10pt,twocolumn]{revtex4}
\usepackage{graphicx,epstopdf}
\usepackage[caption=false]{subfig}
\usepackage{multirow}
\usepackage{appendix}
\usepackage{xcolor}
\usepackage[colorlinks=true, urlcolor=blue, linkcolor=blue, citecolor=green]{hyperref}
\def\pslash{p\!\!\!\slash }
\def\qslash{q\!\!\!\slash }
\def\xslash{x\!\!\!\slash }

\def\beq{\begin{equation}}
\def\eeq{\end{equation}}
\def\bea{\begin{eqnarray}}
\def\eea{\end{eqnarray}}
\def\beeq{\begin{eqnarray}}
\def\eeeq{\end{eqnarray}}

\def\vel{\left|}
\def\ver{\right|}
\def\nnb{\nonumber}

\def\nnb{\nonumber}

\def\ba{\begin{array}}
\def\ea{\end{array}}

\def\xis0{{\Xi^{*0}}}

\def\g5{\gamma_5}

\def\es{\!\!\! &=& \!\!\!}

\def\ar{&+& \!\!\!}
\def\ek{&-& \!\!\!}

\begin{document}

\title{Gravitational transition form factors of $N(1535) \rightarrow N$ }

\author{
 U. \"{O}zdem$^{1,3,*}$
     and    
K. Azizi$^{2,3,\dag}$ 
\\
  \small$^1$ Health Services Vocational School of Higher Education, Istanbul Aydin University, Sefakoy-Kucukcekmece, 34295 Istanbul, Turkey\\
 \small $^2$ Department of Physics, University of Tehran, North Karegar Avenue, Tehran 14395-547, Iran\\
\small $^3$ Department of Physics, Dogus University, Acibadem-Kadikoy, 34722 Istanbul, Turkey\\
$^{*}$e-mail:ulasozdem@aydin.edu.tr\\
$^{\dag}$e-mail:kazem.azizi@ut.ac.ir
}​

\begin{abstract}
We employ the quark part of the symmetric energy-momentum tensor current to calculate the transition gravitational form factors of the $N(1535) \rightarrow N$ by means of the light cone QCD sum rule formalism. 
In numerical analysis, we use two different sets of the shape parameters in the distribution amplitudes of the $ N(1535) $ baryon and the general form of the nucleon's interpolating current.  It is seen that the momentum squared dependence of the gravitational form factors can be well described by the p-pole fit function.
 The results obtained by using two sets of parameters are found to be quite different from each other and  the  $N(1535) \rightarrow N$ transition gravitational form factors depend highly on the shape parameters of the distribution amplitudes of the $N(1535)$ state that parametrize the relative orbital angular momentum of the constituent quarks. 

\end{abstract}
 \date{\today}
\maketitle

\section{Introduction} 
The form factors (FFs) are essential parameters for gaining knowledge on the internal organization of the  composit particles at low energies.
Using various FFs, one can get important information on many quantities such as size, shape, radius, electrical and magnetic charge distributions, axial and tensor charges and other mechanical and electromagnetical parameters of hadrons.
The gravitational FFs (GFFs) or energy-momentum tensor FFs (EMTFFs)  are described by the help of the matrix elements of the symmetric energy-momentum tensor.
Just, as the Fourier transform of the electromagnetic form factors can be explicated with regard to the spatial distribution of electrical charge and magnetization, the Fourier transform of the gravitational form factors can be explicated with regard to the spatial distribution of momentum, energy, pressure etc.
Hence, the investigation of these FFs attracts significant interest for comprehension of the internal structure of the nucleon.
The GFFs of the nucleon have been examined within different theoretical models such as, chiral quark soliton model ($\chi$QSM) \cite{Petrov:1998kf, Schweitzer:2002nm, Ossmann:2004bp, Wakamatsu:2005vk, Wakamatsu:2006dy, Goeke:2007fq,Goeke:2007fp, Jung:2013bya,Jung:2014jja, Jung:2015piw,Wakamatsu:2007uc}, 
lattice QCD \cite{Hagler:2003jd,mathur:1999uf,  Gockeler:2003jfa, Bratt:2010jn, Hagler:2007xi,Brommel:2007sb,Negele:2004iu,Deka:2013zha}, 
 light-cone QCD sum rules (LCSR)~\cite{Anikin:2019kwi,Azizi:2019ytx}, Skyrme model \cite{Cebulla:2007ei,Kim:2012ts}, chiral perturbation theory ($\chi$PT) \cite{chen:2001pva, Belitsky:2002jp, Ando:2006sk, Diehl:2006ya, Diehl:2006js, Dorati:2007bk}, Bag model~\cite{Neubelt:2019sou},  
 and, instant and front forms (IFF) \cite{Lorce:2018egm}.
 Interested readers can find more details about these studies in a recent review~\cite{Polyakov:2018zvc}.

 In this study, we extend our previous work on the nucleon's EMTFFs \cite{Azizi:2019ytx} and  calculate for the first time (to our best knowledge)  the transitional gravitational form factors of $N(1535) \rightarrow N$ due to the energy-momentum tensor current  in the framework of the light-cone QCD sum rule \cite{Braun:1988qv, Balitsky:1989ry, Chernyak:1990ag} using the general form of the interpolating current of nucleon and distribution amplitudes of $N(1535)$ (Hereafter we shall represent $N(1535)$ particle as $N^*$).
 The main advantage of this method is that it is an analytical method and includes direct QCD parameters. The method involves a two-stage approach. First, the corresponding  correlation function is calculated in terms of the quark-gluon properties. Second, it is obtained in terms of the  hadron properties such as GFFs. When calculating the quark-gluon properties, a connection is established between the low energy processes and the QCD vacuum, which is expressed in terms of distribution amplitudes. In this approach, the hadrons are represented by interpolating currents carrying the same quantum number as the hadrons. These interpolating currents are inserted into the correlation function and the short (perturbative) and long distance (nonperturbative) interactions are separated using the operator product expansion (OPE). The hadronic and OPE representations of the same correlation function are then matched. To suppress the unwanted contributions coming from the higher states and continuum,  the Borel transformation and continuum subtraction supplied by the quark-hadron duality assumption are applied. By choosing some independent Lorentz structures from both sides of the resultant equations the desired sum rules for the FFs are obtained in terms of hadronic parameters as well as QCD degrees of freedom.
 %
 The light cone QCD sum rule approach has been successfully applied to calculate different form factors of hadrons as well as the transition form factors among baryons due to the electromagnetic, axial and tensor currents  at   high $Q^2$  (see e.g.~\cite{Aliev:2004ju, Aliev:2007qu, Wang:2006su, Braun:2006hz, Erkol:2011iw, Erkol:2011qh, 
Aliev:2011ku, kucukarslan:2016xhx, Kucukarslan:2015urd, Kucukarslan:2014bla, Kucukarslan:2014mfa,Aliev:2019tmk,Aliev:2008cs}).
%

The presentation of the manuscript is organized as follows. In section \ref{secII}, we briefly discuss the formalism  and calculate the  LCSR for the transition
gravitational form factors under investigation. 
In section \ref{secIII}, we numerically analyze the  $N^* \rightarrow N$ transition GFFs. We fix the auxiliary parameters entering the calculations according to the standard prescriptions of the method. We also use the wave functions and all the including parameters recently available for the DAs of the $ N^*  $ state to find the $ Q^2 $ behavior of the form factors in this section. 
Section \ref{secIV} is dedicated to a discussion of the results achieved for the GFFs.
The  distribution amplitudes of the $N^*$ state and explicit expressions of the $N^* \rightarrow N$ transition GFFs  are presented in the appendixes.

 \section{Formalism}\label{secII}

The following correlation function is used in the analytical calculations to compute the GFFs of the $ N^* \rightarrow N $ transition by means of the LCSR:

\begin{equation}\label{corf}
	\Pi_{\mu\nu}(p,q)=i\int d^4 x e^{iqx} \langle 0 |\mathcal{T}[J_{N}(0)T_{\mu \nu}^q(x)]|N^*(p)\rangle,
\end{equation}
where 
 $J_{N}(0)$ is interpolating current of the nucleon,  the $T_{\mu\nu}^q(x)$ is the quark part of the energy-momentum tensor current
  and $\mathcal{T}$ is the time ordering operator. 

 The main goal in this section is to calculate the above-mentioned correlation function in 
 two different languages and apply the  prescriptions previously discussed. 
 Thus, the calculations of the  hadronic and OPE (QCD) representations of the correlation function are in order.

 \subsection{Hadronic representation of the correlation function}
 
We begin to compute the correlator with regards to the hadronic degrees of freedom with the inclusion of the physical
features of the hadrons under examination. To that end, we embed intermediate states of nucleon with the same quantum number of $J_N(0)$ into the correlator. As a result, we get
 
 \begin{align}\label{phys}
 \Pi_{\mu\nu}^{Had}(p,q) &=\sum_{s{'}} \frac{\langle0|J_N|{N(p',s')}\rangle\langle {N(p',s')}
 |T_{\mu \nu}^q|N^*(p,s)\rangle}{m^2_{N}-p'^2} \nonumber\\
 &+...,
\end{align}
where the contributions coming from the higher states and continuum are denoted by dots.
The expression shown in Eq. \eqref{phys} can be  more simplified by introducing the following matrix elements \cite{Polyakov:2018zvc}:
\begin{widetext}

 \begin{align}\label{GFFs}
 \langle0|J_N|{N(p',s')}\rangle &= \lambda_N u_N(p',s'),\\
 \nonumber\\
 \langle N(p^\prime,s')|T_{\mu\nu}^q|N^*(p,s)\rangle &=
 \bar{u}_N(p^\prime,s')\bigg[A^{N^*-N}(Q^2)\frac{ \tilde P_\mu \tilde P_\nu}{2(m_{N^*}+m_N)}
 +~i J^{N^*-N}(Q^2)\frac{(\tilde P_\mu \sigma_{\nu\rho}+\tilde P_\nu \sigma_{\mu\rho})\Delta^\rho}{2(m_{N^*}+m_N)}
    \nonumber\\
    \nonumber\\
  & +~D^{N^*-N}(Q^2) \frac{\Delta_\mu \Delta_\nu- g_{\mu\nu} \Delta^2}{2(m_{N^*}+m_N)}
  +~ \bar c^{N^*-N} (Q^2) \frac{m_{N^*}+m_N}{2} g_{\mu \nu} \bigg] \gamma_5 u_{N^*}(p,s), 
 \end{align}
 
\end{widetext}
where $\lambda_{N}$ is residue of the nucleon,
$\tilde P= p'+p$, 
$\Delta = p'-p$,  
$\sigma_{\mu\nu}= \frac{i}{2}[\gamma_\mu, \gamma_\nu]$ and  $Q^2=- \Delta^2$.
Here, $A^{N^*-N}(Q^2)  $, $J^{N^*-N}(Q^2)  $, $ D^{N^*-N}(Q^2) $ and $\bar c ^{N^*-N} (Q^2) $ are the GFFs. Summation over spin of nucleon is carried out by
\begin{eqnarray}\label{sspin}
     \sum _{s'} u_N(p',s')\, \bar u_N(p',s') &=& \pslash'+m_N.
\end{eqnarray}

Substituting Eqs. \eqref{GFFs}-\eqref{sspin} in Eq. \eqref{phys}, we achieve the hadronic representation of the correlation function with respect to the hadronic parameters as
\begin{widetext}

 \begin{align}\label{physson}
\Pi_{\mu\nu}^{Had} (p,q)&= \frac{\lambda_N}{m^2_{N}-p'^2}( \pslash'+m_N)\bigg[A^{N^*-N}(Q^2)\frac{ \tilde P_\mu \tilde P_\nu}{2(m_{N^*}+m_N)}
 +~i J^{N^*-N}(Q^2)\frac{(\tilde P_\mu \sigma_{\nu\rho}+\tilde P_\nu \sigma_{\mu\rho})\Delta^\rho}{2(m_{N^*}+m_N)}
    \nonumber\\
    \nonumber\\
  & +~D^{N^*-N}(Q^2) \frac{\Delta_\mu \Delta_\nu- g_{\mu\nu} \Delta^2}{2(m_{N^*}+m_N)}
  +~ \bar c^{N^*-N} (Q^2) \frac{m_{N^*}+m_N}{2} g_{\mu \nu} \bigg] \gamma_5 u_{N^*}(p,s).
\end{align}

\end{widetext}

From the Eq. $\eqref{physson}$, we can decompose the hadronic side of the correlator in terms of  different invariant functions and independent Lorentz structures:

\begin{align}
 \Pi_{\mu\nu}^{Had} (p,q) &=\Pi_1^{Had}(Q^2)\, p^{\prime}_\mu p^{\prime}_\nu \gamma_5 + \Pi_2^{Had}(Q^2)\,p^{\prime}_\mu p^{\prime}_\nu \qslash \gamma_5 \nonumber\\
 &+ \Pi_3^{Had}(Q^2)\,g_{\mu\nu}\qslash \gamma_5 + \Pi_4^{Had}(Q^2)\,g_{\mu\nu} \gamma_5 \nonumber\\
 &+....
\end{align}

\subsection{QCD representation of the correlation function}

To achieve the expression of the correlator in terms of the QCD parameters, the explicit forms for the interpolating currents of the $ J_N(0)$ and $T_{\mu\nu}^q (x)$ are needed.
 These currents are defined by the following expressions with respect to quark fields:
 
\begin{align}
\label{intpol}
 J_N(0) &= 2\,\epsilon^{abc}\bigg[\big[u^{aT}(0) C  d^b(0)\big]\gamma_5 u^c(0)\nonumber\\
& + t\,\big[u^{aT}(0) C \gamma_5  d^b(0)\big] u^c(0)\bigg],\nonumber\\
 T_{\mu\nu}^q (x) &= \frac{i}{2}\bigg[\bar{u}^d(x)\overleftrightarrow{D}_\mu(x) \gamma_\nu u^d(x) \nonumber\\
& + \bar{d}^e(x)\overleftrightarrow{D}_\mu(x) \gamma_\nu d^e(x) 
+(\mu \leftrightarrow \nu) \bigg],
\end{align}
where
the charge conjugation operator, arbitrary mixing parameter; and color indices are denoted as $C$, $t$; and  $a$, $b$, $c$, $d$, $e$, respectively.
The two-sided covariant derivative is defined as
\begin{align}
 \overleftrightarrow{D}_\mu(x) &=\frac{1}{2} \Big[ \overrightarrow{D}_\mu(x) - \overleftarrow{D}_\mu(x) \Big],
\end{align}
with 
\begin{align}
 \overrightarrow{D}_\mu(x) &= \overrightarrow{\partial}_\mu(x)+igA_\mu(x), \\
\overleftarrow{D}_\mu(x) &= \overleftarrow{\partial}_\mu(x) -igA_\mu(x), 
\end{align}
 where $A_\mu$ is the gluon field.
 We neglect the gluon fields contributions, i.e the gluonic part of the EMT 
since considering these contributions needs information of quark-gluon mixed DAs of the
$N^*$ state which unluckily are not available.
Therefore, in this work, we will take notice of
the quark part of the energy-momentum tensor current in Eq. (\ref{intpol}).
 We insert the interpolating currents $ J_N(0)$ and $T_{\mu\nu}^q (x)$ into the 
 correlator and carry out the necessary contractions by the help of the Wick theorem. 
 Consequently, we obtain
 
 \begin{widetext}
 
\begin{align}\label{corrfunc}
	\Pi_{\mu\nu}^{QCD}(p,q)&=-\int d^4 x e^{iqx}\Bigg[\bigg\{ (\gamma_5)_{\gamma\delta}\, C_{\alpha\beta}\, ( \overleftrightarrow{D}_\mu (x)\gamma_\nu)_{\omega \rho} 
	+
	t\, (I)_{\gamma\delta}\,(C \gamma_5)_{\alpha\beta}\,( \overleftrightarrow{D}_\mu(x) \gamma_\nu)_{\omega \rho}\nonumber\\
	&+
	(\gamma_5)_{\gamma\delta} \, C_{\alpha\beta} \, ( \overleftrightarrow{D}_\nu (x)\gamma_\mu)_{\omega \rho} 
    +
	t\,(I)_{\gamma\delta}\,(C \gamma_5)_{\alpha\beta}\, ( \overleftrightarrow{D}_\nu (x)\gamma_\mu)_{\omega \rho}
	\bigg\}\nonumber\\
   & \times	
      \bigg\{\Big (\delta_\sigma^\alpha \delta_\theta^\rho \delta_\phi^\beta S_u(-x)_{\delta \omega}
     +\, \delta_\sigma^\delta \delta_\theta^\rho \delta_\phi^\beta S_u(-x)_{\alpha \omega}\Big)
     \langle 0|\epsilon^{abc} u_{\sigma}^a(0) u_{\theta}^b(x) d_{\phi}^c(0)|N^*(p)\rangle \nonumber\\
     &+\delta_\sigma^\alpha \delta_\theta^\delta \delta_\phi^\rho S_d(-x)_{\beta \omega} 
    \,\langle 0|\epsilon^{abc} u_{\sigma}^a(0) u_{\theta}^b(0) d_{\phi}^c(x)|N^*(p)\rangle\bigg\}
   \Bigg],
\end{align}

\end{widetext}
where $S_q(x)$ is propagator of the light $q= u, d $ quarks and it is identified as

\begin{align}
\label{edmn09}
S_{q}(x) &= 
\frac{1}{2 \pi^2 x^2}\Big( i \frac{{\xslash}}{x^{2}}-\frac{m_{q}}{2 } \Big)
- \frac{\langle  \bar qq \rangle }{12} \Big(1-i\frac{m_{q} \xslash}{4}   \Big)\nonumber\\
&- \frac{\langle \bar q \sigma.G q \rangle }{192}x^2  \Big(1-i\frac{m_{q} \xslash}{6}   \Big)
\nonumber\\
&-\frac {i g_s }{32 \pi^2 x^2} ~G^{\mu \nu} (x) \bigg[\rlap/{x}
\sigma_{\mu \nu} +  \sigma_{\mu \nu} \rlap/{x}
 \bigg].
\end{align}
It should be noted here that we work on $m_q = 0$ limit, 
therefore the terms proportional with quark mass do not give any contribution. 
The terms proportional with quark ($\langle  \bar qq \rangle$) and mixed ($\langle \bar q \sigma.G q \rangle$) condensates are killed by performing Borel transformation.
%
The contributions coming from the terms corresponding  to the gluon strength field tensor ($G_{\mu\nu}$) are expected to be small \cite{Diehl:1998kh}, 
 which are relevant to the distribution amplitudes of four- and five-particles.
Therefore, we can neglect the contributions of these terms in calculations as well.
%
As a result, just the first term of the light quark propagator contributes to our computations.
The $\langle 0| \epsilon^{abc} u_{\sigma}^a(a_1 x) u_{\theta}^b(a_2 x) d_{\phi}^c(a_3 x)|N^*(p)\rangle$ 
matrix element is the expression containing the distribution amplitudes of the $N^*$ state and it is required for further calculations. The  explicit form of this matrix element in terms of the related  DAs  together with the explicit forms of DAs  for $N^*$ are given in the Appendix A. Using the distribution amplitudes of $N^*$ state and  applying the integration over x, 
the QCD representation of the correlation function is acquired as

\begin{align}
 \Pi_{\mu\nu}^{QCD} (p,q)&=\Pi_1^{QCD}(Q^2)\, p^{\prime}_\mu p^{\prime}_\nu \gamma_5 + \Pi_2^{QCD}(Q^2)\,p^{\prime}_\mu p^{\prime}_\nu \qslash \gamma_5 \nonumber\\
& + \Pi_3^{QCD}(Q^2)\,g_{\mu\nu}\qslash \gamma_5 + \Pi_4^{QCD}(Q^2)\,g_{\mu\nu} \gamma_5 \nonumber\\
&+.... 
\end{align}
The explicit expressions of the $\Pi_1^{QCD}(Q^2)$, $\Pi_2^{QCD}(Q^2)$, $\Pi_3^{QCD}(Q^2)$ and $\Pi_4^{QCD}(Q^2)$  are  presented in the Appendix B.

\subsection{Light-cone QCD sum rules for the \texorpdfstring{$ N^{*} \rightarrow N$}{} transition }
The required light-cone QCD sum rules for the $N^* \rightarrow N $ transition GFFs are obtained by matching the coefficients of 
different Lorentz structures from both the hadronic and QCD representations of the correlator.  

We shall note that we employ the structures $p^{\prime}_\mu p^{\prime}_\nu \gamma_5$, $p^{\prime}_\mu p^{\prime}_\nu \qslash \gamma_5$, 
$q_\mu q_\nu \qslash \gamma_5$ and $g_{\mu \nu} \gamma_5$ 
to find the light-cone QCD sum rules for  the transition GFFs, $A^{N^*-N}(Q^2)$, $J^{N^*-N}(Q^2)$, $D^{N^*-N}(Q^2)$ and $\bar c^{N^*-N}(Q^2)$, respectively.
 Hence, 

\begin{align}
 \frac{\lambda_N }{{m^2_{N}-p'^2}}\,A^{N^*-N}(Q^2)   &=- \frac{m_N+m_{N^*}}{2 m_{N^*}} \varPi_1^{QCD}(Q^2),\\
 \frac{\lambda_N }{{m^2_{N}-p'^2}}\, J^{N^*-N}(Q^2)  &= \frac{m_N+m_{N^*}}{2} \varPi_2^{QCD}(Q^2),\\
 \frac{\lambda_N }{{m^2_{N}-p'^2}}\, D^{N^*-N}(Q^2)  &= -2(m_N+m_{N^*}) \varPi_3^{QCD}(Q^2),\\
\frac{\lambda_N }{{m^2_{N}-p'^2}}\, \bar c^{N^*-N} (Q^2)  &= - \frac{2}{m^2_{N^*}}\varPi_4^{QCD}(Q^2).
\end{align}


For the calculation of the $N^* \rightarrow N$ transition GFFs the residue of nucleon, $\lambda_N$ is needed, as well. 
The residue of the nucleon is determined from
two point sum rules \cite{Aliev:2011ku}:
\begin{align}
\label{residue}
\lambda_N^2 e^{-\frac{m_N^2}{M^2}} &=~\frac{M^6}{256 \pi^4} (5+2 t + t^2) E_2(z)  
- \frac{\langle \bar{q}q \rangle^2}{6}  \nonumber\\
& \times \bigg\{ 6 (1-t^2)
-(1-t)^2   - \frac{m_0^2 }{4 M^2} \Big[12 (1-t^2) \nonumber\\
&- (1-t)^2  \Big] \bigg\}, 
\end{align}
where
\begin{eqnarray}
\label{nolabel}
z &=& s_0/M^2,\nonumber 
\end{eqnarray}
and 
\begin{eqnarray}
E_n(z)&=&1-e^{-z}\sum_{i=0}^{n}\frac{z^i}{i!}~. \nonumber
\end{eqnarray}

\section{Numerical Results}\label{secIII}

This section is dedicated to the numerical analysis of the $N^* \rightarrow N$ transition GFFs.
To this end we need distribution amplitudes of $N^*$ state. 
The explicit expressions of these distribution amplitudes are given in the Appendix A. 
For further calculations we need the shape parameters of the distribution amplitudes of the $N^*$ state, which are presented in Table \ref{tab:data}. Additionally, we use: $m_N = 0.94$ GeV, $m_{N^*} = 1.51 \pm 0.01$ GeV~\cite{Tanabashi:2018oca},  
 $m_q = 0$, $\langle \bar{q}q\rangle=(-0.24\pm 0.01)^3$~GeV$^3$  and  $m_0^2=0.8 \pm 0.1$~GeV$^2$~\cite{Ioffe:2005ym}.  
 
 \begin{widetext}

   \begin{table}[htp]
\caption{Numerical values of the shape parameters of the distribution amplitudes for the $N^*$ state at renormalization scale $\mu^2 =
	2.0~\mathrm{GeV}^2$. Besides these values, we use  $ \lambda_1^N m_N = -3.88 (2)(19)\times 10^{-2}$ GeV$^3$ and  $\lambda_2^{N^*}  m_{N^*} = 8.97 (45)\times 10^{-2}$ GeV$^3$, given in Ref. \cite{Braun:2014wpa} at renormalization $\mu^2 = 4.0$ GeV$^2$, by rescaling to  $\mu^2 =
	2.0~\mathrm{GeV}^2$. }\label{tab:data}
\renewcommand{\arraystretch}{1.3}
\addtolength{\arraycolsep}{-0.5pt}
\small
$$
\begin{array}{|c|c|c|c|c|c|c|c|c|c|}
\hline \hline
\mbox{Model} &\mid \lambda_1^{N^\ast}/\lambda_1^N \mid &
f_{N^\ast}/\lambda_1^{N^\ast} & \varphi_{10} & \varphi_{11} & \varphi_{20} &
\varphi_{21} & \varphi_{22} & \eta_{10} & \eta_{11} \\  \hline
\mbox{ LCSR--1} & 0.633 & 0.027 & 0.36 & -0.95 & 0 & 0 & 0 & 0     & 0.94 \\
\mbox{LCSR--2}  & 0.633 & 0.027 & 0.37 & -0.96 & 0 & 0 & 0 & -0.29 & 0.23 \\
\hline \hline
\end{array}
$$
\renewcommand{\arraystretch}{1}
\addtolength{\arraycolsep}{-1.0pt}
\end{table}

\end{widetext}

The LCSR for the $N^* \rightarrow N$ transition GFFs also include some auxiliary parameters: the mixing parameter $t$, the continuum threshold $s_0$ and the 
Borel mass parameter square $M^2$. The GFFs should not be affected by the changes of these parameters much. Hence, we search for working windows for these auxiliary parameters
such that in these working windows the GFFs relatively weakly depend on these parameters. 
The mixing parameter $t$ is chosen such that, 
the estimation of the GFFs is independent of the value of $t$ in its working region. 
From the numerical calculations, it is obtained that in the region -0.2 $\leq cos\theta \leq$ -0.45  the GFFs weakly depend on $t$, where   tan$\theta$ = $t$. 
The working region for continuum threshold $s_0$ is acquired taking into account the fact that the GFFs are almost insensitive with respect to its changes, as well. We choose the $s_0$ in the interval  $2.5$ GeV$^ 2$ $\leq s_0 \leq 3.0$ GeV$^ 2 $.
We use the following steps to achieve the working interval for the Borel mass parameter $ M^2 $.
The lower cutoff of $M^2$  is obtained demanding that the perturbative part exceeds the nonperturbative one and the series of nonperturbative terms are convergent.
The upper cutoff of $M^2$ is acquired  by the condition that the contributions of higher states and continuum should be less than the ground states contribution.
 These requirements are both fulfilled when $M^2$ varies in the interval $2.0$ GeV$^ 2$ $\leq M^2 \leq 3.5$ GeV$^ 2 $.


In Figs. (\ref{Msqfigs})-(\ref{Qsqfigs}), we plot the dependence of the GFFs $A^{N^*-N}$($Q^2$),  $J^{N^*-N}$($Q^2$), $D^{N^*-N}$($Q^2$) and $\bar c^{N^*-N}$($Q^2$) on different parameters for the various values 
of the arbitrary mixing parameter $t$ and other parameters in their working regions for two sets of the distribution amplitudes for the $N^*$ baryon.
 As it can be seen from Figs. (\ref{Msqfigs}) and (\ref{s0sqfigs}), 
the obtained values of the GFFs are approximately independent of the  continuum threshold $s_0$ and  Borel mass parameter $M^2$  on their working regions.
Therefore, for numerical calculations of the central values of GFFs we will use the central values of the $M^2$ and $s_0$.
   The LCSR method is reliable only $Q^2 > 1.0$ GeV$^2$. However, the mass corrections of the distribution amplitudes $\sim m^2_{N^*}/Q^2$ become quite large for $Q^2 < 2.0$ GeV$^2$ namely the LCSR turn out to be unreliable.
  Therefore, for GFFs, we expect the LCSR to operate efficiently and effectively in the $2.0 $ GeV$^2 \leq Q^2  \leq 6.0$ GeV$^2$ region.
    In Fig. \ref{Qsqfigs}, we present the $Q^2$ dependency of the GFFs on the fixed Borel mass parameter and continuum threshold and various values of the arbitrary mixing parameter $t$. 
  We see that GFFs smoothly vary in terms of $Q^2$ as expected. 
  We also observe that the GFFs $A^{N^*-N}$($Q^2$),  $J^{N^*-N}$($Q^2$) and $D^{N^*-N}$($Q^2$) are sensitive to the variations of the mixing parameter $t$ for both sets.
  %
  %
  The $\bar c^{N^*-N}$($Q^2$) form factor is obtained using only  the first set of the distribution amplitudes. 
  The second set of the distribution amplitudes gives unphysical results  for this form factor (the form factor changes its sign in the region under consideration), therefore it is not presented.

To extend the behavior of GFFs to the region $0 \leq Q^2 < 2$ we need to use some fit parametrizations.
Our numerical calculations show that the GFFs of the $N^* \rightarrow N$ transition can be properly characterized using the p-pole fit function
 \begin{align}
{\cal F}(Q^2)= \frac{{\cal F}(0)}{\Big(1+ Q^2/(p\,m^2_{p})\Big)^p}.
\end{align} 

Our results for the fit parameters of the  $N^* \rightarrow N$ transition GFFs are presented in Table \ref{fit_table}.
The results, obtained by using two sets, are  found to be quite different from each other.
As it can be see from the Table \ref{tab:data}, the main difference between two sets of the distribution amplitudes is the numerical values for the shape parameters $\eta_{10}$ and $\eta_{11}$, 
which are connected to the three quark wave functions of the p-wave $N^*$ baryon. 
 As a result we find that the  $N^* \rightarrow N$ transition GFFs depend strongly to the input parameters of the distribution amplitudes of the $N^*$ baryon that parametrize the  relative orbital angular momentum of the quarks. 
%
There are no theoretical predictions for the transition GFFs under study in the literature to be compared with our results.
The  $D(Q^2)$ GFF of the $N^*$ has recently been  obtained by means of the Bag model,  $D(Q^2=0) = -12.97$ \cite{Neubelt:2019sou}.
 Any future experimental data will help us  gain useful knowledge on the DAs of the $N^*$ state and its nature and internal structure.
%
%
%
%
\begin{widetext}

\begin{table}[htp]
\caption{ The obtained numerical results for the parameters of the GFFs by using the p-pole fit functions.
}
\hspace*{-0.5cm}
\begin{tabular}{ |l|c|c|c|c|c|c|}
\hline\hline
&\multicolumn{3}{|c|}{LCSR-1} &\multicolumn{3}{|c|}{LCSR-2}\\
\hline\hline
\begin{tabular}{c}Form Factors \end{tabular}& \begin{tabular}{c} ${\cal F}(0)$ \end{tabular} & \begin{tabular}{c}$m_{p}$(GeV) \end{tabular}& \begin{tabular}{c}p \end{tabular}
&${\cal F}(0)$ & $m_{p}$(GeV) & p \\ \hline\hline
        $A^{N^*-N}$($Q^2$)      &$~~1.33\pm 0.13$  &$1.30 \pm 0.10 $ &$3.0 - 3.4$ & $~~0.63 \pm 0.10$ &$1.32 \pm 0.10$ &$3.0 - 3.4$ \\
        $J^{N^*-N}$($Q^2$)      &$~~0.27 \pm 0.07$ & $1.13 \pm 0.10 $ &$3.0 - 3.4$ & $~~0.13 \pm 0.05$ & $1.17 \pm 0.11$ &$2.9 - 3.3$ \\
        $D^{N^*-N}$($Q^2$)      &$- 8.20 \pm 2.02$& $1.14 \pm 0.10 $ & $3.6 - 4.0 $ & $-6.91 \pm 1.80$ &$ 1.02 \pm 0.09$ &$3.2 - 3.6$  \\
       $\bar c^{N^*-N}$($Q^2$)  &$-0.40\pm 0.06$ & $1.18 \pm 0.10$ &$3.0 - 3.5$&$-$ &$-$ &$-$\\
\hline \hline
\end{tabular}
	\label{fit_table}
\end{table}

\end{widetext}

\section{Summary and Concluding Remarks}\label{secIV}
We applied the quark part of the symmetric energy-momentum tensor current to compute, for the first time, the  GFFs of the $ N(1535) \rightarrow N $ transition with the help of the LCSR approach. 
Studying the GFFs of the particles give valuable knowledge about the total angular momentum, spatial distribution of energy,  pressure and shear forces inside the particles, etc.
In numerical analysis, we used two different sets of shape parameters in the distribution amplitudes of the $ N(1535)$ state and took into account the most general form of the nucleon's interpolating current.
 It is seen that the momentum squared dependence of the gravitational form factors can be well described by a p-pole fit function.
 The results obtained by using two sets have been found to be quite different from each other.
  We  found that the values  of the  $N(1535) \rightarrow N$ transition GFFs highly depend on the input parameters of 
  the distribution amplitudes of the $N(1535)$ state that parameterize relative orbital angular momentum of the quarks.
 As  previously mentioned, we calculated the $ N(1535) \rightarrow N $ transition GFFs for the first time in the literature. 
 Therefore, there are no experimental data or theoretical predictions to be compared with  the results obtained in this study. Calculations of the  GFFs from different methods and approaches are of great importance.
 Such calculations and comparison of the obtained results with each other will not only  help us get useful information on the DAs of the $N^*$ state, but also  experimental groups for measuring the values of the related GFFs.
 
 \section{Acknowledgments}
U. \"{O}. acknowledges the support of  the
Scientific and Technological Research Council of Turkey
(TUBITAK) provided  through the 2218-National Postdoctoral Research Scholarship Program.



\begin{widetext}
\section*{Appendix A: Distribution amplitudes of $N^\ast$ state
}

In this appendix, we present the explicit forms of $N^*$ DAs \cite{maxiphd}:
\bea\label{wave func}
&& 4 \langle 0 \vel \epsilon^{abc} u_\alpha^a(a_1 x) d_\beta^b(a_2 x)
d_\gamma^c(a_3 x) \ver N^\ast(p)\rangle\nnb\\
\es \mathcal{S}_1 m_{N^\ast}C_{\alpha\beta}N_{\gamma}^\ast -
\mathcal{S}_2 m_{N^\ast}^2 C_{\alpha\beta}(\rlap/x N^\ast)_{\gamma}\nnb\\
\ar \mathcal{P}_1 m_{N^\ast}(\gamma_5 C)_{\alpha\beta}(\gamma_5 N^\ast)_{\gamma} +
\mathcal{P}_2 m_{N^\ast}^2 (\gamma_5 C)_{\alpha\beta}(\gamma_5 \rlap/x N^\ast)_{\gamma} -
\left(\mathcal{V}_1 + \frac{x^2 m_{N^\ast}^2}{4} \mathcal{V}_1^M \right) 
(\rlap/p C)_{\alpha\beta} N_{\gamma}^\ast \nnb\\
\ar \mathcal{V}_2 m_{N^\ast}(\rlap/p C)_{\alpha\beta}(\rlap/x N^\ast)_{\gamma} + 
\mathcal{V}_3 m_{N^\ast}(\gamma_\mu C)_{\alpha\beta} (\gamma^\mu N^\ast)_{\gamma} -
\mathcal{V}_4 m_{N^\ast}^2 (\rlap/x C)_{\alpha\beta} N_{\gamma}^\ast \nnb\\
\ek \mathcal{V}_5 m_{N^\ast}^2(\gamma_\mu C)_{\alpha\beta}
(i \sigma^{\mu\nu} x_\nu N^\ast)_\gamma +
\mathcal{V}_6 m_{N^\ast}^3 (\rlap/x C)_{\alpha\beta}(\rlap/x N^\ast)_{\gamma} \nnb \\
\ek \left(\mathcal{A}_1 + \frac{x^2m_{N^\ast}^2}{4}\mathcal{A}_1^M\right)
(\rlap/p \gamma_5 C)_{\alpha\beta} (\gamma N^\ast)_{\gamma} +
\mathcal{A}_2 m_{N^\ast}(\rlap/p \gamma_5 C)_{\alpha\beta} (\rlap/x \gamma_5 N^\ast)_{\gamma} +
\mathcal{A}_3 m_{N^\ast}(\gamma_\mu\gamma_5 C)_{\alpha\beta}
(\gamma^\mu \gamma_5 N^\ast)_{\gamma}\nnb\\
\ek \mathcal{A}_4 m_{N^\ast}^2(\rlap/x \gamma_5 C)_{\alpha\beta}
(\gamma_5 N^\ast)_{\gamma} -
\mathcal{A}_5 m_{N^\ast}^2(\gamma_\mu \gamma_5 C)_{\alpha\beta}
(i \sigma^{\mu\nu} x_\nu \gamma_5 N^\ast)_{\gamma} +
\mathcal{A}_6 m_{N^\ast}^3(\rlap/x \gamma_5 C)_{\alpha\beta}
(\rlap/x \gamma_5 N^\ast)_{\gamma}\nnb\\
\ek \left(\mathcal{T}_1 + \frac{x^2m_{N^\ast}^2}{4}\mathcal{T}_1^M \right)
(i \sigma_{\mu\nu} p_\nu C)_{\alpha\beta} (\gamma^\mu N^\ast)_{\gamma} +
\mathcal{T}_2 m_{N^\ast} (i \sigma_{\mu\nu} x^\mu p^\nu C)_{\alpha\beta}
N_{\gamma}^\ast\nnb\\
\ar \mathcal{T}_3 m_{N^\ast}(\sigma_{\mu\nu} C)_{\alpha\beta}
(\sigma^{\mu\nu} N^\ast)_{\gamma} +
\mathcal{T}_4 m_{N^\ast} (\sigma_{\mu\nu} p^\nu C)_{\alpha\beta}
(\sigma^{\mu\rho} x_\rho N^\ast)_{\gamma} \nnb\\
\ek \mathcal{T}_5 m_{N^\ast}^2 (i\sigma_{\mu\nu} x^\nu C)_{\alpha\beta}
(\gamma^\mu N^\ast)_{\gamma} -  
\mathcal{T}_6 m_{N^\ast}^2 (i \sigma_{\mu\nu} x^\mu p^\nu C)_{\alpha\beta}
(\rlap/x N^\ast)_{\gamma}\nnb\\
\ek \mathcal{T}_7 m_{N^\ast}^2 (\sigma_{\mu\nu} C)_{\alpha\beta}
(\sigma^{\mu\nu} \rlap/x N^\ast)_{\gamma} +
\mathcal{T}_8 m_{N^\ast}^3 (\sigma_{\mu\nu} x^\nu C)_{\alpha\beta}
(\sigma^{\mu\rho} x_\rho N^\ast)_{\gamma}~.\nnb
\eea
The “calligraphic” functions can be represented with respect to the functions of the specific twists as:
  \begin{flalign*}
		   \mathcal{S}_1 =& S_1,\hspace{3.5cm}                    2p.x\mathcal{S}_2=S_1-S_2,\nonumber\\
		   \mathcal{P}_1=&P_1, \hspace{3.5cm}                     2p.x\mathcal{P}_2=P_1-P_2,\\
           \mathcal{V}_1=&V_1,\hspace{3.5cm}                      2p.x\mathcal{V}_2=V_1-V_2-V_3, \nonumber\\
           2\mathcal{V}_3=&V_3,\hspace{3.5cm}                     4p.x\mathcal{V}_4=-2V_1+V_3+V_4+2V_5,\nonumber\\
           4p.x\mathcal{V}_5=&V_4-V_3,\hspace{2.5cm}              4(p.x)^2\mathcal{V}_6=-V_1+V_2+V_3+V_4 + V_5-V_6\\
           \mathcal{A}_1=&A_1, \hspace{3.5cm}                     2p.x\mathcal{A}_2=-A_1+A_2-A_3,\nonumber\\
        2\mathcal{A}_3=&A_3,\hspace{3.5cm}                        4p.x\mathcal{A}_4=-2A_1-A_3-A_4+2A_5, \nonumber\\
        4p.x\mathcal{A}_5=&A_3-A_4, \hspace{2.5cm}                4(p.x)^2\mathcal{A}_6=A_1-A_2+A_3+A_4-A_5+A_6\\
            \mathcal{T}_1=&T_1, \hspace{3.5cm}                    2p.x\mathcal{T}_2=T_1+T_2-2T_3, \nonumber\\
            2\mathcal{T}_3=&T_7,\hspace{3.5cm}                    2p.x\mathcal{T}_4=T_1-T_2-2T_7,\nonumber\\
        2p.x\mathcal{T}_5=&-T_1+T_5+2T_8,\hspace{1cm}             4(p.x)^2\mathcal{T}_6=2T_2-2T_3-2T_4+2T_5+2T_7+2T_8,        \nonumber\\ 
        4p.x \mathcal{T}_7=&T_7-T_8,\hspace{2.5cm}                4(p.x)^2\mathcal{T}_8=-T_1+T_2 +T_5-T_6+2T_7+2T_8,
\end{flalign*}
where $V_i, A_i, T_i,S_i $ and $P_i$  are vector, axialvector, tensor, scalar and pesudoscalar  distribution amplitudes, respectively.
The explicit expressions of these functions are given as follows
	\begin{eqnarray*}
		V_1(x_i,\mu)&=&120x_1x_2x_3[\phi_3^0(\mu)+\phi_3^+(\mu)(1-3x_3)],\nonumber\\
		V_2(x_i,\mu)&=&24x_1x_2[\phi_4^0(\mu)+\phi_4^+(\mu)(1-5x_3)],\nonumber\\
		V_3(x_i,\mu)&=&12x_3\{\psi_4^0(\mu)(1-x_3)+\psi_4^-(\mu)[x_1^2+x_2^2-x_3(1-x_3)]
		\nonumber\\&&+\psi_4^+(\mu)(1-x_3-10x_1x_2)\},\nonumber\\
		V_4(x_i,\mu)&=&3\{\psi_5^0(\mu)(1-x_3)+\psi_5^-(\mu)[2x_1x_2-x_3(1-x_3)]
	        \nonumber\\&&+\psi_5^+(\mu)[1-x_3-2(x_1^2+x_2^2)]\},\nonumber\\
	        V_5(x_i,\mu)&=&6x_3[\phi_5^0(\mu)+\phi_5^+(\mu)(1-2x_3)],\nonumber\\
		V_6(x_i,\mu)&=&2[\phi_6^0(\mu)+\phi_6^+(\mu)(1-3x_3)],\nonumber\\
                A_1(x_i,\mu)&=&120x_1x_2x_3\phi_3^-(\mu)(x_2-x_1),\nonumber\\
                A_2(x_i,\mu)&=&24x_1x_2\phi_4^-(\mu)(x_2-x_1),\nonumber\\
		A_3(x_i,\mu)&=&12x_3(x_2-x_1)\{(\psi_4^0(\mu)+\psi_4^+(\mu))+\psi_4^-(\mu)(1-2x_3)
		\},\nonumber\\
		A_4(x_i,\mu)&=&3(x_2-x_1)\{-\psi_5^0(\mu)+\psi_5^+(\mu)(1-2x_3)+\psi_5^-(\mu)x_3\},\nonumber\\
		A_5(x_i,\mu)&=&6x_3(x_2-x_1)\phi_5^-(\mu)\nonumber\\
		A_6(x_i,\mu)&=&2(x_2-x_1)\phi_6^-(\mu),\nonumber\\
		T_1(x_i,\mu)&=&120x_1x_2x_3[\phi_3^0(\mu)+\frac{1}{2}(\phi_3^--\phi_3^+)(\mu)(1-3x_3)],\nonumber\\
		T_2(x_i,\mu)&=&24x_1x_2[\xi_4^0(\mu)+\xi_4^+(\mu)(1-5x_3)],\nonumber\\
		T_3(x_i,\mu)&=&6x_3\{(\xi_4^0+\phi_4^0+\psi_4^0)(\mu)(1-x_3)+
		(\xi_4^-+\phi_4^--\psi_4^-)(\mu)[x_1^2+x_2^2-x_3\\
		&&(1-x_3)]+(\xi_4^++\phi_4^++\psi_4^+)(\mu)(1-x_3-10x_1x_2)\},\nonumber\\
		T_4(x_i,\mu)&=&\frac{3}{2}\{(\xi_5^0+\phi_5^0+\psi_5^0)(\mu)(1-x_3)+
		(\xi_5^-+\phi_5^--\psi_5^-)(\mu)[2x_1x_2-x_3(1-x_3)]\nonumber\\
		&&+(\xi_5^++\phi_5^++\psi_5^+)(\mu)(1-x_3-2(x_1^2+x_2^2))\},\nonumber\\
		T_5(x_i,\mu)&=&6x_3[\xi_5^0(\mu)+\xi_5^+(\mu)(1-2x_3)],\nonumber\\
		T_6(x_i,\mu)&=&2[\phi_6^0(\mu)+\frac{1}{2}(\phi_6^--\phi_6^+)(\mu)(1-3x_3)],\nonumber \\
		T_7(x_i,\mu)&=&6x_3\{(-\xi_4^0+\phi_4^0+\psi_4^0)(\mu)(1-x_3)+ (-\xi_4^-+\phi_4^--\psi_4^-)(\mu)[x_1^2+x_2^2
		\nonumber\\&&-x_3(1-x_3)](1-x_3)]
		+(-\xi_4^++\phi_4^++\psi_4^+)(\mu)(1-x_3-10x_1x_2)\},\nonumber\\
		T_8(x_i,\mu)&=&\frac{3}{2}\{(-\xi_5^0+\phi_5^0+\psi_5^0)(\mu)(1-x_3)+
		(-\xi_5^-+\phi_5^--\psi_5^-)(\mu)[2x_1x_2- \\&&+x_3(1-x_3)](1-x_3)]
		 +(-\xi_5^++\phi_5^++\psi_5^+)(\mu)(1-x_3-2(x_1^2+x_2^2))\},\nonumber\\
		S_1(x_i,\mu) &=& 6 x_3 (x_2-x_1) \left[ (\xi_4^0 + \phi_4^0 +
		\psi_4^0 + \xi_4^+ + \phi_4^+ + \psi_4^+)(\mu) \right.\\
		&&\left.+ (\xi_4^- + \phi_4^- - \psi_4^-)(\mu)(1-2 x_3) \right]\nonumber \\
		S_2(x_i,\mu) &=& \frac{3}{2} (x_2 -x_1) \left[- \left(\psi_5^0 +
		\phi_5^0 + \xi_5^0\right)(\mu) + \left(\xi_5^- + \phi_5^- - \psi_5^-
		\right)(\mu) x_3 \right. \nonumber \\
		&& \left.+\left(\xi_5^+ + \phi_5^+ + \psi_5^0 \right)(\mu) (1- 2x_3)\right]\nonumber \\
					P_1(x_i,\mu) &=& 6 x_3 (x_2-x_1) \left[ (\xi_4^0 - \phi_4^0 -
		\psi_4^0 + \xi_4^+ - \phi_4^+ - \psi_4^+)(\mu)\right. \\
		&&\left. + (\xi_4^- - \phi_4^-+ \psi_4^-)(\mu)(1-2 x_3) \right]\nonumber \\
		P_2(x_i,\mu) &=& \frac32 (x_2 -x_1) \left[\left(\psi_5^0 + \psi_5^0
		- \xi_5^0\right)(\mu) + \left(\xi_5^- - \phi_5^- + \psi_5^-\right)(\mu) x_3 \right. \nonumber\\
		&& \left. + \left(\xi_5^+ - \phi_5^+ - \psi_5^+ \right)(\mu) (1- 2x_3)\right]\, .\\
		{\cal V}_1^M (x_2) &=& \int \limits_0^{1-x_2} dx_1 V_1^{M}(x_1, x_2,1-x_1-x_2)= \frac{x_2^2}{24} \left[ f_{N^\ast} C_{f}^u (x_2) + 
\lambda_1^{N^\ast} C_{\lambda}^u (x_2)\right],\nnb\\
{\cal A}_1^{M} (x_2) &=& \int_0^{1-x_2} dx_1 A_1^{M}(x_1, x_2,1-x_1-x_2)= \frac{ x_2^2}{24} (1 - x_2)^3 \left[ f_{N^\ast} D_{f}^u (x_2)  + 
\lambda_1^{N^\ast} D_{\lambda}^u(x_2) \right],\nnb \\
{\cal T}_1^{M} (x_2) &=& \int_0^{1-x_2} dx_1 T_1^{M}(x_1, x_2,1-x_1-x_2) = \frac{ x^2}{48}  \left[ f_{N^\ast} E_{f}^u(x_2)  +
\lambda_1^{N^\ast} E_{\lambda}^u(x_2)  \right].\nnb
	      \end{eqnarray*}

The subsequent functions  come across to the above DAs and they can be parametrized with respect to the independent parameters, such as  $f_{N^\ast}$,
$\lambda_1$, $\lambda_2$, $f_1^u$, $f_1^d$, $f_2^d$, $A_1^u$ and $V_1^d$:\\

\begin{align}
\label{DAs}
&\phi_3^0 = \phi_6^0 = f_{N^\ast} \nnb \\
&\phi_4^0 = \phi_5^0 = {1 \over 2} (f_{N^\ast} + \lambda_1^{N^\ast}) \nnb \\
&\xi_4^0 = \xi_5^0 = {1 \over 6} \lambda_2^{N^\ast} \nnb \\
&\psi_4^0 = \psi_5^0 = {1 \over 2} (f_{N^\ast} - \lambda_1^{N^\ast}), \nnb\\
&\phi_3^- = {21 \over 2} f_{N^\ast} A_1^u,~~\phi_3^+ = {7 \over 2}
f_{N^\ast} (1 - V_1^d), \nnb \\
&\phi_4^+ = {1\over 4} \left[f_{N^\ast} (3-10 V_1^d) + \lambda_1^{N^\ast}
(3 - 10 f_1^d) \right], \nnb \\
&\phi_4^- = - {5\over 4} \left[f_{N^\ast} (1-2 A_1^u) - \lambda_1^{N^\ast}
(1 - 2 f_1^d - 4 f_1^u) \right], \nnb \\
&\psi_4^+ = - {1\over 4} \left[f_{N^\ast} (2 + 5 A_1^u - 5 V_1^d) - 
\lambda_1^{N^\ast} (2 - 5 f_1^d - 5 f_1^u) \right], \nnb \\
&\psi_4^- = {5\over 4} \left[f_{N^\ast} (2 - A_1^u - 3 V_1^d) - 
\lambda_1^{N^\ast} (2 - 7 f_1^d + f_1^u) \right], \nnb \\
&\xi_4^+ = {1\over 16} \lambda_2^{N^\ast} (4 - 15 f_2^d), \nnb \\
&\xi_4^- = {5\over 16} \lambda_2^{N^\ast} (4 - 15 f_2^d), \nnb \\
&\phi_5^+ =  - {5\over 6} \left[f_{N^\ast} (3 + 4 V_1^d) - \lambda_1^{N^\ast}
(1 - 4 f_1^d)\right], \nnb \\
&\phi_5^- = - {5\over 3} \left[f_{N^\ast} (1 - 2 A_1^u) -
\lambda_1^{N^\ast} (f_1^d - f_1^u)\right], \nnb \\
&\psi_5^+ = - {5\over 6} \left[f_{N^\ast} (5 + 2 A_1^u - 2 V_1^d) - 
\lambda_1^{N^\ast} (1 - 2 f_1^d - 2 f_1^u) \right],\nnb \\
&\psi_5^- = {5\over 3} \left[f_{N^\ast} (2 - A_1^u - 3 V_1^d) +     
\lambda_1^{N^\ast} (f_1^d - f_1^u) \right], \nnb \\
&\xi_5^+ = {5\over 36} \lambda_2^{N^\ast}  (2 - 9 f_2^d)~,\nnb \\
&\xi_5^- = -{5\over 4} \lambda_2^{N^\ast} f_2^d, \nnb \\
&\phi_6^+ =  {1\over 2} \left[f_{N^\ast} (1 - 4 V_1^d) -
\lambda_1^{N^\ast} (1 - 2 f_1^d)\right], \nnb \\
&\phi_6^- = {1\over 2} \left[f_{N^\ast} (1 + 4 A_1^d) +                   
\lambda_1^{N^\ast} (1 - 4 f_1^d - 2 f_1^u)\right], \nnb\\
&C_{f}^u (x_2) =  (1 - x_2)^3 \Big[113 + 495x_2 - 552x_2^2 
- 10A_1^u(1 - 3x_2) +  2V_1^d(113 - 951x_2 + 828x_2^2) \Big]~,\nnb \\
&C_{\lambda}^u (x_2)  =  - (1- x_2)^3 
              \Big[13 - 20f_1^d + 3x_2 + 10f_1^u(1 \!-\! 3x_2)\Big]~. \nnb\\
&D_{f}^u(x_2)  =  11 + 45 x_2 - 2 A_1^u (113 - 951 x_2 + 828 x_2^2 )
+ 10 V_1^d (1 - 30 x_2)~, \nnb \\
& D_{\lambda}^u (x_2) =   29 - 45 x_2 - 10 f_1^u (7 - 9 x_2) - 
20 f_1^d (5 - 6 x_2)~. \nnb\\
&E_{f}^u (x_2)  =  -\Big\{
(1 - x_2) \Big[3 (439 + 71 x_2 - 621 x_2^2 + 587 x_2^3 
- 184 x_2^4) + 4 A_1^u (1 - x_2)^2 (59 - 483 x_2 + 414 x_2^2) \nnb \\
&- 4 V_1^d (1301 - 619 x_2 - 769 x_2^2 + 1161 x_2^3 - 414 x_2^4)\Big]
\Big\}
- 12 (73 - 220V_1^d) \ln[x_2]~, \nnb \\
&E_{\lambda}^u (x_2)  =  -\Big\{(1 - x_2) 
\Big[5 - 211 x_2 + 281 x_2^2 - 111 x_2^3 
+ 10 (1 + 61 x_2 - 83 x_2^2 + 33 x_2^3) f_1^d \nnb \\
&- 40(1 - x_2)^2 (2 - 3 x_2) f_1^u\Big]
\Big\} - 12 (3 - 10 f_1^d) \ln[x_2],\nnb
\end{align}

where the parameters $A_1^u,~V_1^d,~f_1^d,~f_1^u$, and $f_2^d$ are defined as
\cite{maxiphd}
\begin{align}
&A_1^u = \varphi_{10} + \varphi_{11}~,\nnb \\
&V_1^d = {1\over 3} - \varphi_{10} + {1\over 3} \varphi_{11}~,\nnb \\
&f_1^u = {1\over 10} - {1\over 6} {f_{N^\ast} \over \lambda_1^{N^\ast}} -
{3\over 5} \eta_{10} - {1\over 3} \eta_{11}~, \nnb \\
&f_1^d = {3\over 10} - {1\over 6} {f_{N^\ast} \over \lambda_1^{N^\ast}} +  
{1\over 5} \eta_{10} - {1\over 3} \eta_{11}~, \nnb \\
&f_2^d = {4\over 15} + {2\over 5} \xi_{10}~.\nnb
\end{align}


\section*{Appendix B: Explicit forms of the functions \texorpdfstring{$\varPi_{i}$}{}  for the \texorpdfstring{$ N^* \rightarrow N$}{} transition }
\begin{align}
\varPi_1^{QCD}(Q^2) &= \frac{m_{N^*}}{2} \int_0^1 \frac{ x_2^2\,dx_2}{(q-p x_2)^2} \int_0^{1-x_2} dx_1\Big[3(1-t)[-A_3-V_3] \, +\,(1+t)[P_1+S_1+T_1-T_7]\Big] (x_1,x_2,1-x_1-x_2)\nonumber\\
&+\frac{m_{N^*}}{2} \int_0^1 \frac{ x_3^2\,dx_3}{(q-p x_3)^2} \int_0^{1-x_3} dx_3\Big[(1-t)[-A_3-V_3] \, +\,(1+t)[P_1+S_1+T_1-T_7]\Big] (x_1,1-x_1-x_3,x_3)\nonumber\\
&+\frac{m^3_{N^*}}{2} \int_0^1 \frac{x_2^2\, dx_2}{(q-p x_2)^4} \int_0^{1-x_2} dx_1\Big[2(1-t)[2A_1^M+V_1^M] \, +\,3\,(1+t)T_1^M\Big] (x_1,x_2,1-x_1-x_2)\nonumber\\
&+ \frac{m_{N^*}}{2}  \int_0^1 \frac{\alpha\, d\alpha}{(q-p \alpha)^2}  \int_\alpha^1 dx_2 \int_0^{1-x_2} dx_1 
\Big[(1-t)[-3A_1+3A_2-3A_3+V_1-V_2-V_3]\nonumber\\
&+ \, (1+t)[5T_1-T_2-3T_3-7T_7]\Big](x_1,x_2,1-x_1-x_2)\nonumber\\
& +\frac{m_{N^*}}{2}  \int_0^1 \frac{\alpha\, d\alpha}{(q-p \alpha)^2}  \int_\alpha^1 dx_3 \int_0^{1-x_3} dx_1 \Big[(1+t)[3T_1+T_2-2T_3-T_7]\Big](x_1,1-x_1-x_3,x_3)\nonumber\\
&+ \frac{m^3_{N^*}}{2}  \int_0^1 \frac{\alpha^3\, d\alpha}{(q-p \alpha)^4}  \int_\alpha^1 dx_2 \int_0^{1-x_2} dx_1 
\Big[(1-t)[-3A_1+A_2-2A_3+A_4+2A_5+V_1-V_2-2V_3+V_4]\nonumber\\
&+ \, (1+t)[P_1-P_2+S_1-S_2+2T_1-3T_2+2T_3-T_5-6T_7]\Big](x_1,x_2,1-x_1-x_2)\nonumber\\
& + \frac{m^3_{N^*}}{2}  \int_0^1 \frac{\alpha^3\, d\alpha}{(q-p \alpha)^4} \int_\alpha^1 dx_3 \int_0^{1-x_3} dx_1 \Big[(1-t)[-A_1+A_2-2A_3+A_4+V_1-V_2-V_3+V_4]\nonumber\\
&\,+\,(1+t)[P_1-P_2+S_1-S_2+2T_1-3T_2+2T_3-T_5-6T_7]\Big](x_1,1-x_1-x_3,x_3) \nonumber\\
 &+ \frac{m^3_{N^*}}{2}  \int_0^1 \frac{\beta^2 \, d\beta}{(q-p \beta)^4} \int_0^\beta d\alpha \int_\alpha^1 dx_2 \int_0^{1-x_2} dx_1
 \Big[2(1-t)[-2A_1+2A_2-2A_3-2A_4+2A_5-2A_6\nonumber\\
 &+V_1-V_2-V_3-V_4-V_5+V_6] \,+\,(1+t)[7T_1-2T_2-5T_3-5T_4-2T_5+7T_6-9T_7\nonumber\\
 &-9T_8]\Big](x_1,x_2,1-x_1-x_2)\nonumber\\
 & + \frac{m^3_{N^*}}{2}  \int_0^1 \frac{\beta^2 \, d\beta}{(q-p \beta)^4} \int_0^\beta d\alpha \int_\alpha^1 dx_3 \int_0^{1-x_3} dx_1
 \Big[(1+t)[2T_1+T_2-3T_3-3T_4+T_5+T_6-T_7\nonumber\\
 &-T_8]\Big](x_1,1-x_1-x_3,x_3),\\
 \nonumber\\
   \varPi_2^{QCD}(Q^2) &=-\frac{1}{2} \int_0^1\frac{1}{(q-px_2)^2} dx_2 \int_0^{1-x_2} dx_1 \Big[(1-t)[A_1+V_1]\Big](x_1,x_2,1-x_1-x_2)\nonumber\\
 & +\frac{m^2_{N^*}}{2} \int_0^1\frac{1}{(q-px_2)^4} dx_2 \int_0^{1-x_2} dx_1 \Big[3(1-t)[A_1^M+V_1^M]\Big](x_1,x_2,1-x_1-x_2)\nonumber
 \end{align}
\begin{align}
&+ \frac{m^2_{N^*}}{2}  \int_0^1 \frac{\alpha^2\, d\alpha}{(q-p \alpha)^4}   \int_\alpha^1 dx_2 \int_0^{1-x_2} dx_1 
\Big[(1-t)[-A_1-A_2-A_3+A_4+A_5-V_1+V_2+V_4]\nonumber\\
&+ \, (1+t)[P_1-P_2+S_1-S_2+T_2-T_5]\Big](x_1,x_2,1-x_1-x_2)\nonumber\\
& +\frac{m^2_{N^*}}{2}  \int_0^1 \frac{\alpha^2\, d\alpha}{(q-p \alpha)^4}  \int_\alpha^1 dx_3 \int_0^{1-x_3} dx_1 \Big[(1-t)[A_1-A_2+A_4-V_1+V_2+V_4]\,+\,(1+t)[P_1-P_2+S_1
 \nonumber\\
&-S_2+T_2-T_5]\Big](x_1,1-x_1-x_3,x_3)\nonumber\\
&+ \frac{m^2_{N^*}}{2}  \int_0^1 \frac{\beta \, d\beta}{(q-p \beta)^4} \int_0^\beta d\alpha  \int_\alpha^1 dx_2 \int_0^{1-x_2} dx_1
 \Big[(1+t)[T_2-T_3-T_4+T_5+T_7+T_8]\Big]\nonumber\\
 &\times(x_1,x_2,1-x_1-x_2)\nonumber\\
 & +\frac{m^2_{N^*}}{2}  \int_0^1 \frac{\beta \, d\beta}{(q-p \beta)^4} \int_0^\beta d\alpha \int_\alpha^1 dx_3 \int_0^{1-x_3} dx_1
 \Big[(1+t)[T_2-T_3-T_4+T_5+T_7+T_8]\Big] \nonumber\\
 &\times(x_1,1-x_1-x_3,x_3),\\
 \nonumber\\
%
%
%
%
%
%
  \varPi_3^{QCD}(Q^2) &=\frac{1}{2} \int_0^1 dx_2 \frac{(1-x_2)}{(q-px_2)^2}\int_0^{1-x_2} \Big[(1-t)[A_1+V_1]\Big](x_1,x_2,1-x_1-x_2)
  \nonumber\\
  &+\frac{1}{2} \int_0^1 dx_3 \frac{(1-x_3)}{(q-px_3)^2}\int_0^{1-x_3} \Big[(1-t)[A_1-V_1]\Big](x_1,1-x_1-x_3,x_3)\nonumber\\
  &+\frac{1}{2} \int_0^1 dx_2 \frac{(1-x_2)}{(q-px_2)^4}\int_0^{1-x_2} \Big[(1-t)[3V_1^M+3A_1^M]\Big](x_1,x_2,1-x_1-x_2)\nonumber\\
 &+ \frac{m^2_{N^*}}{2}  \int_0^1 \frac{d\alpha}{(q-p \alpha)^4}   \int_\alpha^1 dx_2 \int_0^{1-x_2} dx_1 \Big[
  (1-t)[-A_1-A_2-2A_3+A_4+2A_5-V_1+V_2+V_4]\nonumber\\
  &+(1+t)[P_1-P_2+S_1-S_2+T_2-T_5]\Big](x_1,x_2,1-x_1-x_2)
 \nonumber\\
  &+ \frac{m^2_{N^*}}{2}  \int_0^1 \frac{d\alpha}{(q-p \alpha)^4}   \int_\alpha^1 dx_3 \int_0^{1-x_3} dx_1 \Big[
  (1-t)[-A_1-A_2+A_4-V_1+V_2+V_4]\nonumber\\
  &+(1+t)[P_1-P_2+S_1-S_2+T_2-T_5]\Big](x_1,1-x_1-x_3,x_3))
  \nonumber\\
  &+ \frac{m^2_{N^*}}{2}  \int_0^1 \frac{\alpha(1-\alpha)}{(q-p \alpha)^4} d\alpha  \int_\alpha^1 dx_2 \int_0^{1-x_2} dx_1 \Big[
  2(1-t)[-A_1-A_2-2A_3+A_4+2A_5-V_1+V_2+V_4]\nonumber\\
  &+(1+t)[P_1-P_2+S_1-S_2+T_2-T_5]\Big](x_1,x_2,1-x_1-x_2)\nonumber\\
 & +\frac{m^2_{N^*}}{2}  \int_0^1 \frac{\alpha(1-\alpha)}{(q-p \alpha)^4} d\alpha \int_\alpha^1 dx_3 \int_0^{1-x_3} dx_1
 \Big[(1-t)[A_1-A_2+A_4-V_1+V_2+V_4]\nonumber\\
 &+\,(1+t)[P_1-P_2+S_1-S_2-T_2-T_5]\Big](x_1,1-x_1-x_3,x_3),\nonumber\\
 &+ \frac{m^2_{N^*}}{2}  \int_0^1 \frac{ (1+\beta)}{(q-p \beta)^4} d\beta \int_0^\beta d\alpha  \int_\alpha^1 dx_2 \int_0^{1-x_2} dx_1
 (1+t)[T_2-T_3-T_4+T_5+ T_7+T_8]\nonumber\\
 &\times(x_1,x_2,1-x_1-x_2)\nonumber\\
 & +\frac{m^2_{N^*}}{2}  \int_0^1 \frac{ (1+\beta)}{(q-p \beta)^4} d\beta \int_0^\beta d\alpha  \int_\alpha^1 dx_3 \int_0^{1-x_3} dx_1
 \Big[3(1+t)[T_2-T_3-T_4+T_5+ T_7+T_8]\Big]\nonumber\\
 &\times(x_1,1-x_1-x_3,x_3),\\
 \nonumber\\
  \varPi_4^{QCD}(Q^2) &= \frac{m^3_{N^*}}{2} \int_0^1 \frac{dx_2}{(p-px_2)^2}\int_0^{1-x_2}(1+t)T_1^M (x_1,x_2,1-x_1-x_2)\nonumber\\
&+\frac{m^3_{N^*}}{8}  \int_0^1 \frac{\alpha\,d\alpha}{(q-p \alpha)^2} \int_\alpha^1 dx_2 \int_0^{1-x_2} dx_1 \Big[
  (1-t)[-2A_1-2A_2-3A_3+A_4+4A_5-2V_1+2V_2-V_3 \nonumber\\
& +3V_4]+\, (1+t)[2P_1-2P_2+2S_1-2S_2-T_1+T_2+4T_3-4T_5-2T_7]\Big] (x_1,x_2,1-x_1-x_2)\nonumber
\end{align}
%
%
\begin{align}
&+\frac{m^3_{N^*}}{8}  \int_0^1 \frac{\alpha\,d\alpha}{(q-p \alpha)^2}  \int_\alpha^1 dx_3 \int_0^{1-x_3} dx_1 \Big[-2A_1+2A_2-3A_3+A_4+2V_1-2V_2-3V_3+V_4] \nonumber\\
& +\, (1+t)[2P_1-2P_2+2S_1-2S_2+T_1-T_2-3T_5-T_7]\Big](x_1,1-x_1-x_3,x_3) \nonumber\\
&+ \frac{m^3_{N^*}}{8}  \int_0^1 \frac{ d\beta}{(q-p \beta)^2} \int_0^\beta d\alpha  \int_\alpha^1 dx_2 \int_0^{1-x_2} dx_1
 \Big[4\,(1-t)[-A_1+A_2-A_3-A_4+A_5-A_6-V_1+V_2+V_3\nonumber\\
 & +V_4+V_5-V_6] \,+\,(1+t)[T_1-T_2-T_5+T_6-2 T_7-2T_8]\Big](x_1,x_2,1-x_1-x_2)\nonumber\\
 & +\frac{m^3_{N^*}}{8}  \int_0^1 \frac{ d\beta}{(q-p \beta)^2} \int_0^\beta d\alpha  \int_\alpha^1 dx_3 \int_0^{1-x_3} dx_1
 \Big[3(1+t)[T_1-T_2-T_5+T_6-2T_7\nonumber\\
 &-2T_8]\Big](x_1,1-x_1-x_3,x_3),
\end{align}
%
%

As we noticed previously, in QCD side, we start the calculations in x-space then transfer them to the momentum space by performing the corresponding fourier integrals.
To eliminate the unwanted contributions coming from  the excited and continuum states in the correlation function, we carry out  the Borel transformation. 
After the Borel transformation, the contribution of the unwanted terms is exponentially suppressed. 
We also apply the continuum subtraction procedure. 
The Borel transformation and continuum subtraction are performed using the subsequent rules \cite{Braun:2006hz}:
\begin{align}
		\int dz \frac{\rho(z)}{\Theta^2}\rightarrow &-\int_{x_0}^1\frac{dz}{z}\rho(z) e^{-s(z)/M^2}, \nonumber		\\
		\int dz \frac{\rho(z)}{\Theta^4}\rightarrow & \frac{1}{M^2} \int_{x_0}^1\frac{dx}{z^2}\rho(z) e^{-s(z)/M^2}
		+\frac{\rho(x_0)}{Q^2+x_0^2 m^2_{N^*}} e^{-s_0/M^2},\nonumber
	\label{subtract3}
\end{align}
where,
\begin{align}
\Theta =& q-zp,\nonumber\\
s(z) =& (1-z)m^2_{N^*}+\frac{1-z}{z}Q^2,\nonumber\\
x_0 =& \Big(\sqrt{(Q^2+s_0-m^2_{N^*})^2+4 m^2_{N^*} Q^2}-(Q^2+s_0-m^2_{N^*})\Big)/{2m^2_{N^*}}.\nonumber
\end{align}

\end{widetext}

 \begin{widetext}
 
 \begin{figure}[htp]
\centering
  \subfloat[]{\includegraphics[width=0.3\textwidth]{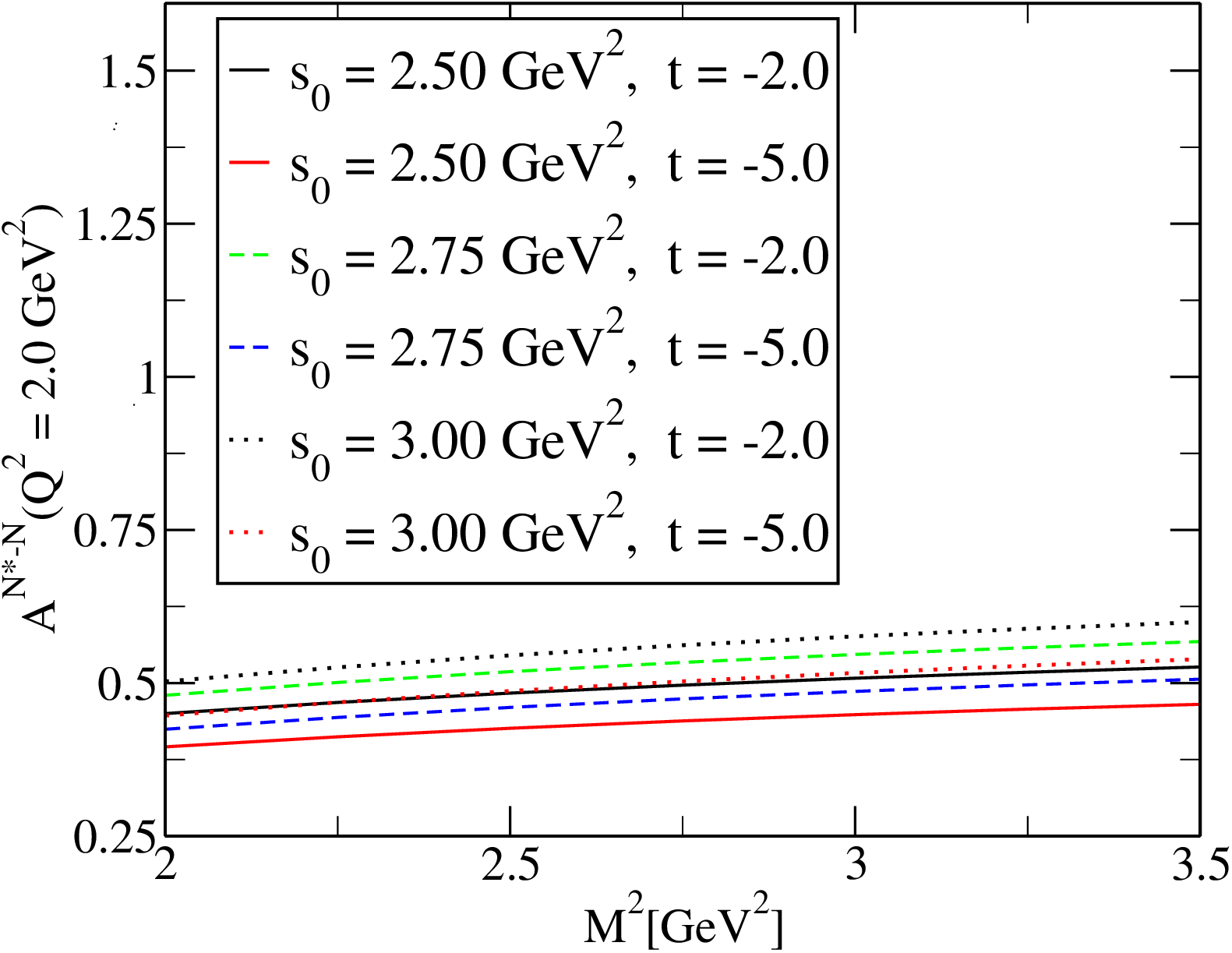}}~~~~~~~~~
   \subfloat[]{\includegraphics[width=0.3\textwidth]{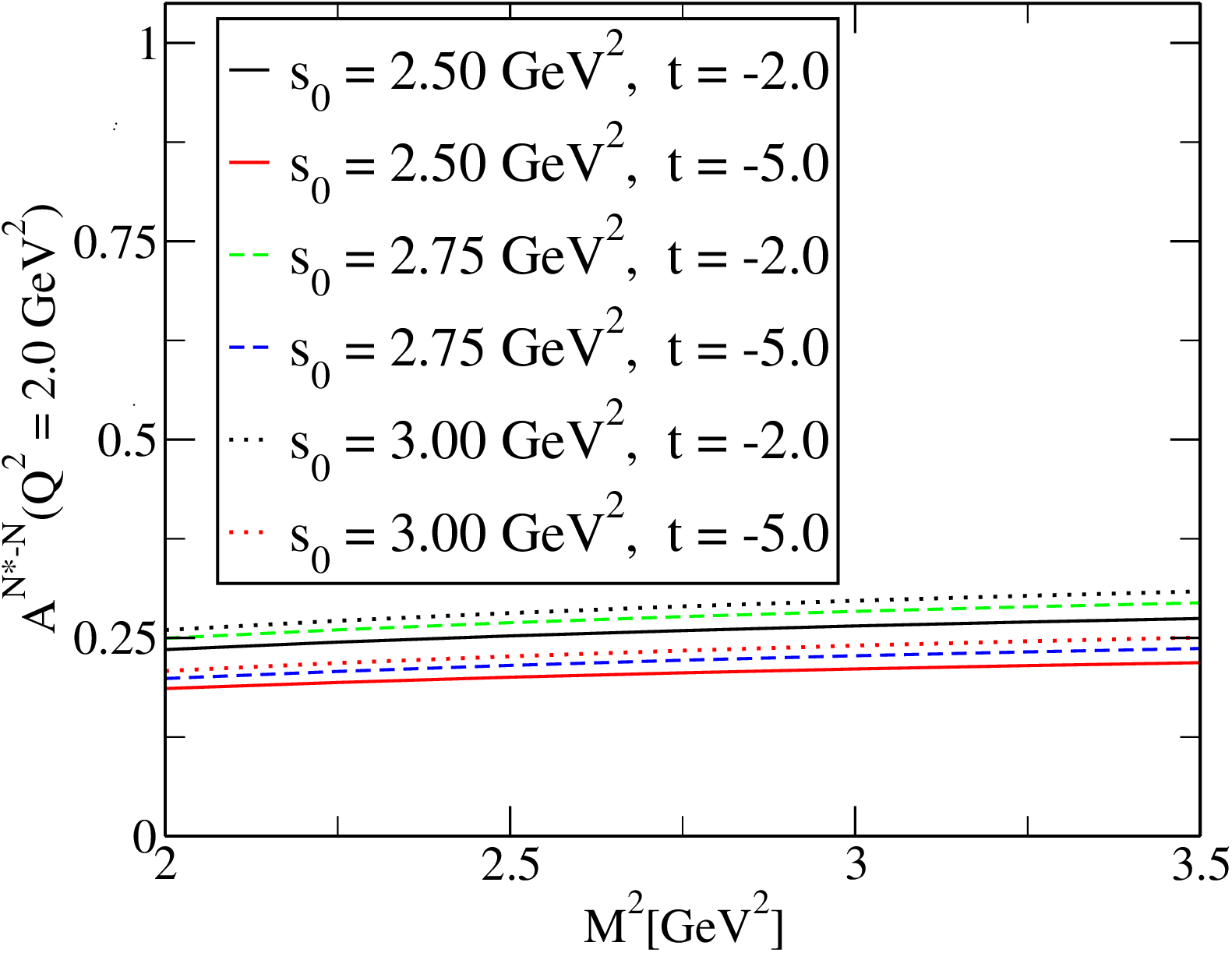}}\\
 \vspace{0.2cm}
 \subfloat[]{\includegraphics[width=0.3\textwidth]{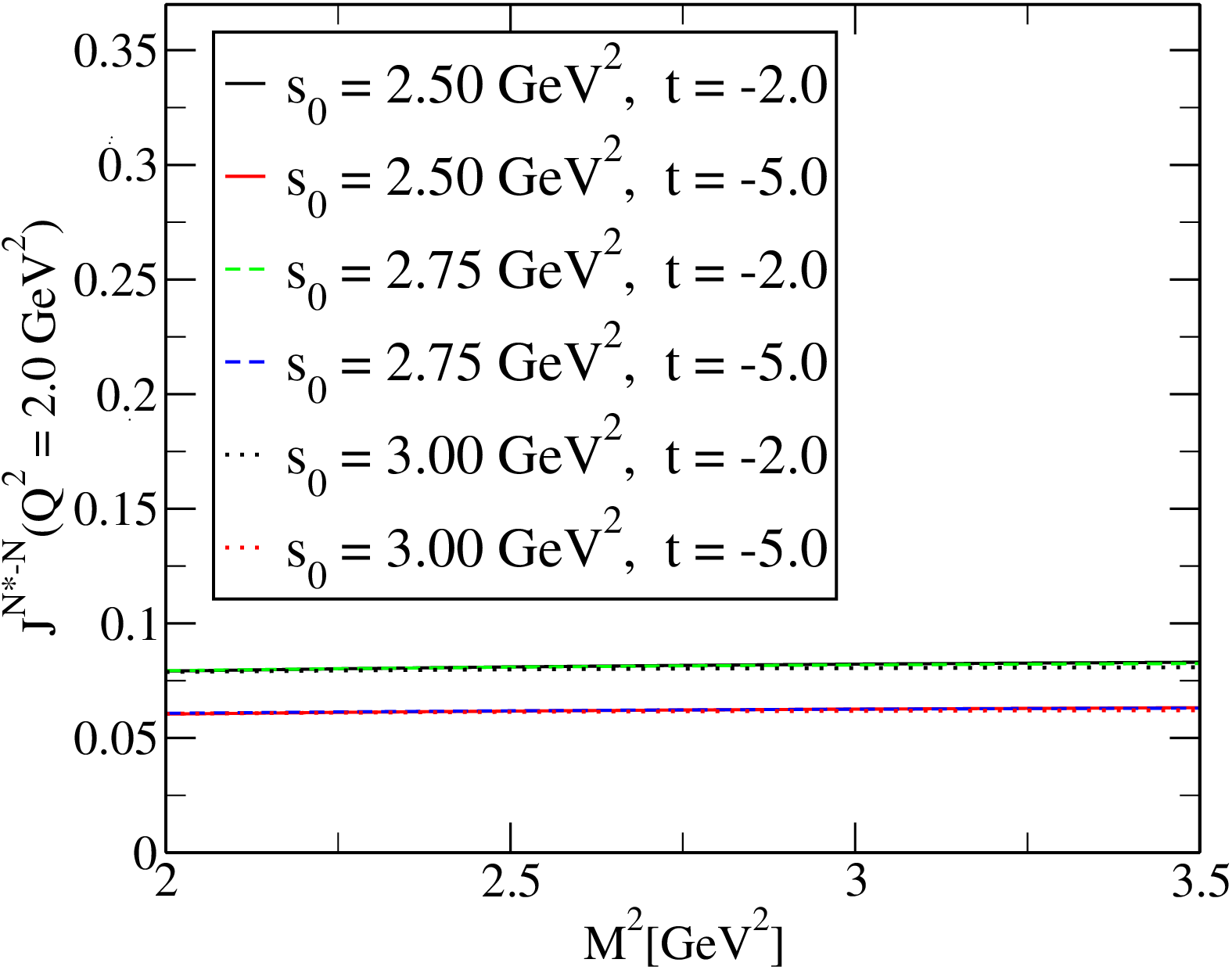}}~~~~~~~~~
 \subfloat[]{\includegraphics[width=0.3\textwidth]{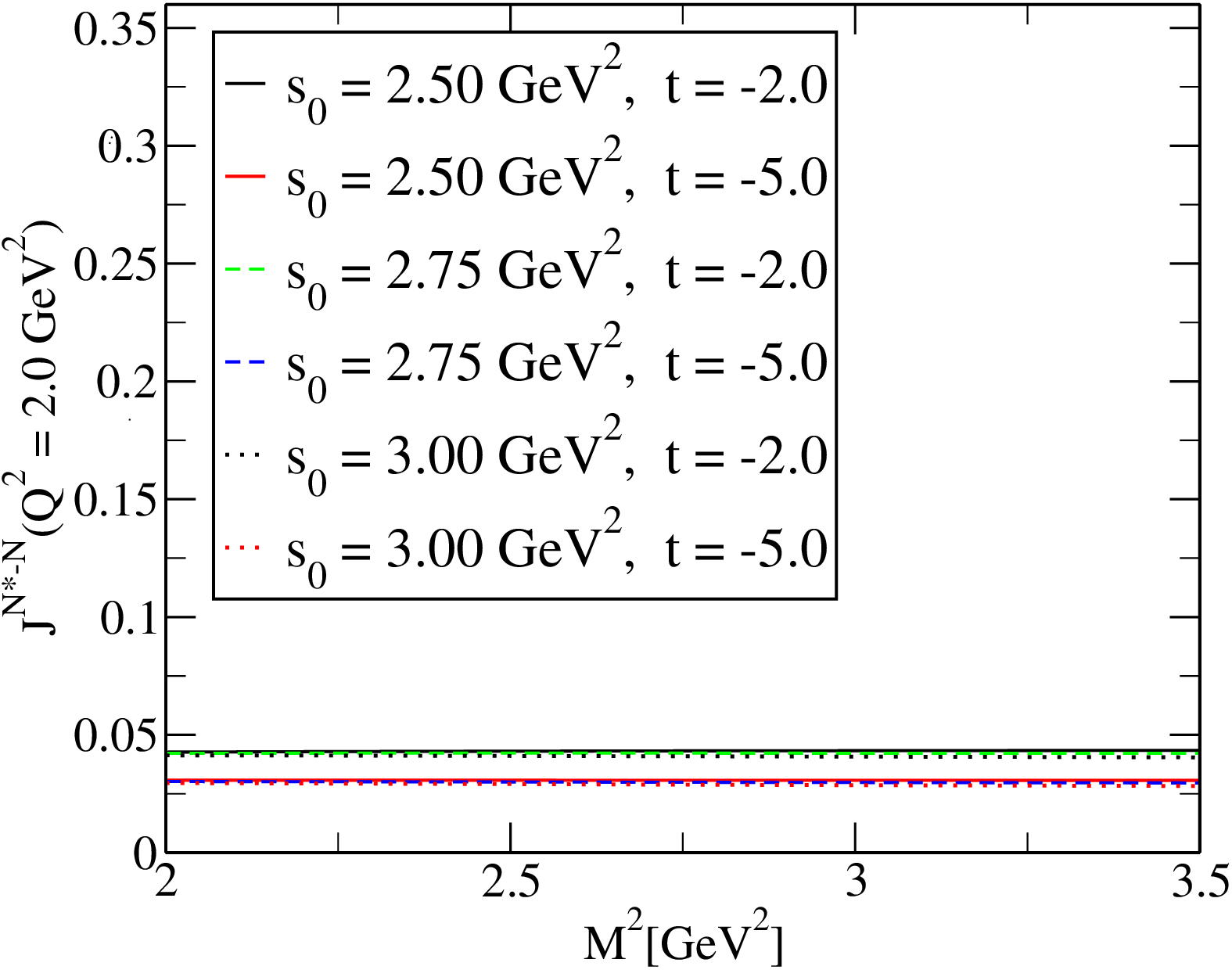}}\\
  \vspace{0.2cm}
  \subfloat[]{\includegraphics[width=0.3\textwidth]{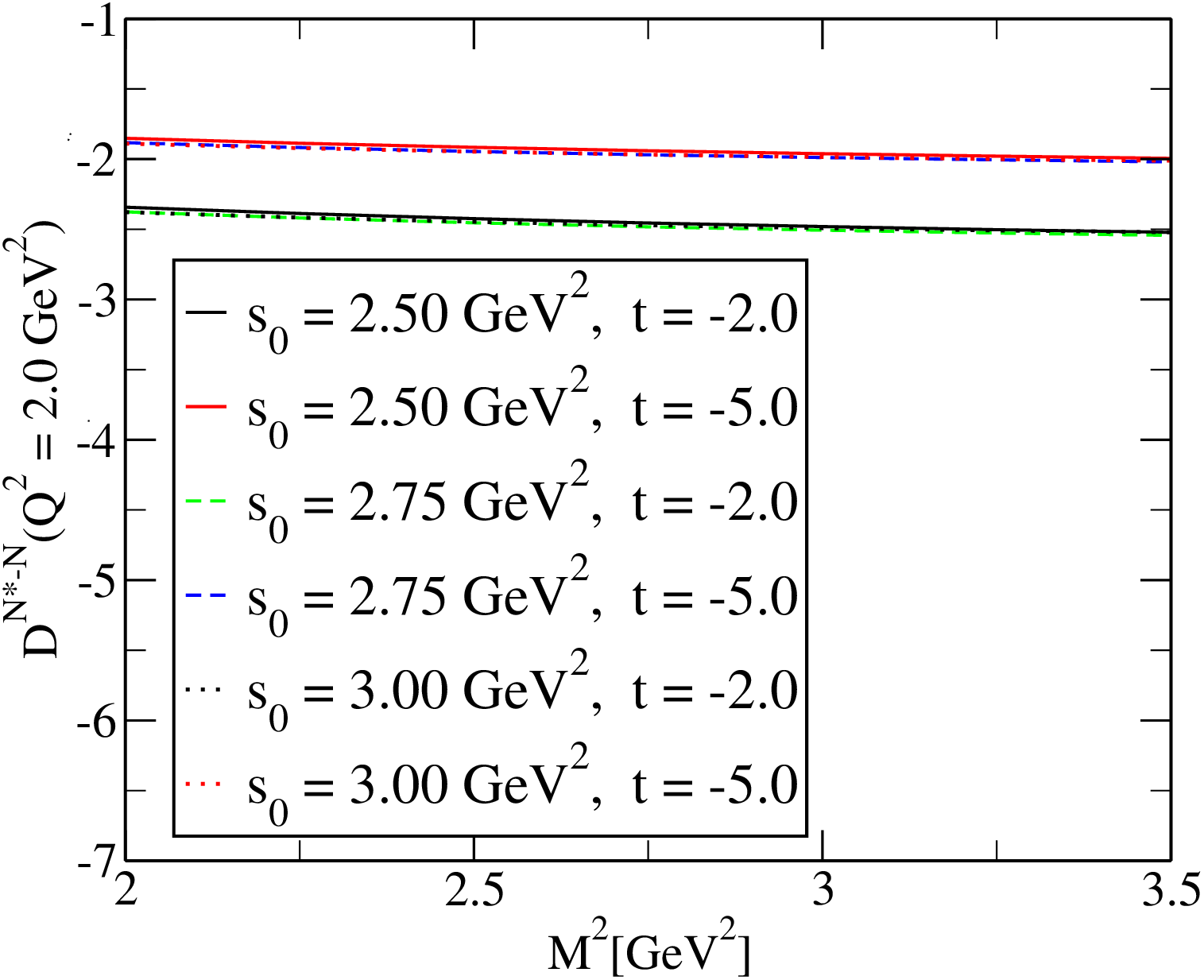}}~~~~~~~~~
   \subfloat[]{\includegraphics[width=0.3\textwidth]{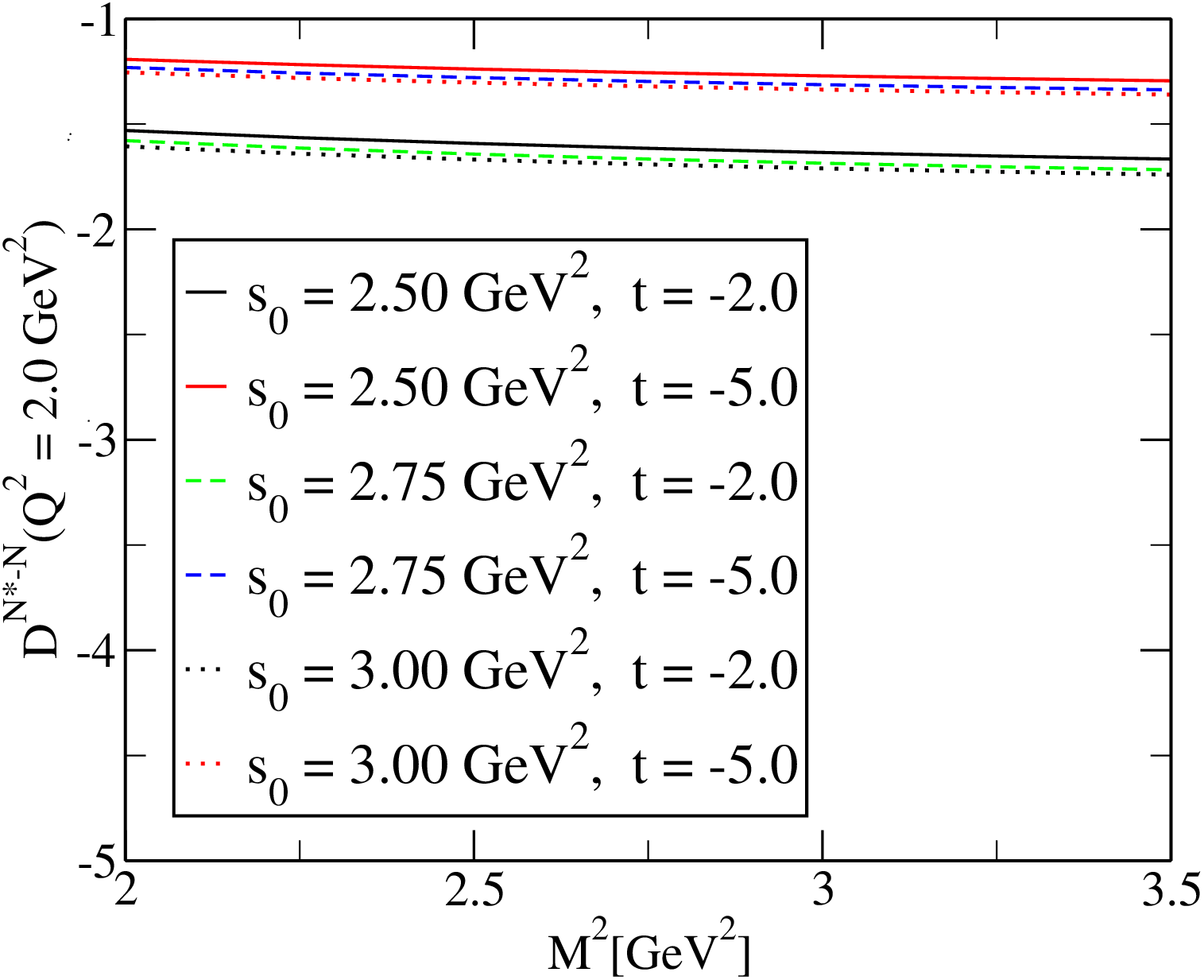}}\\
  \vspace{0.2cm}
  \subfloat[]{\includegraphics[width=0.3\textwidth]{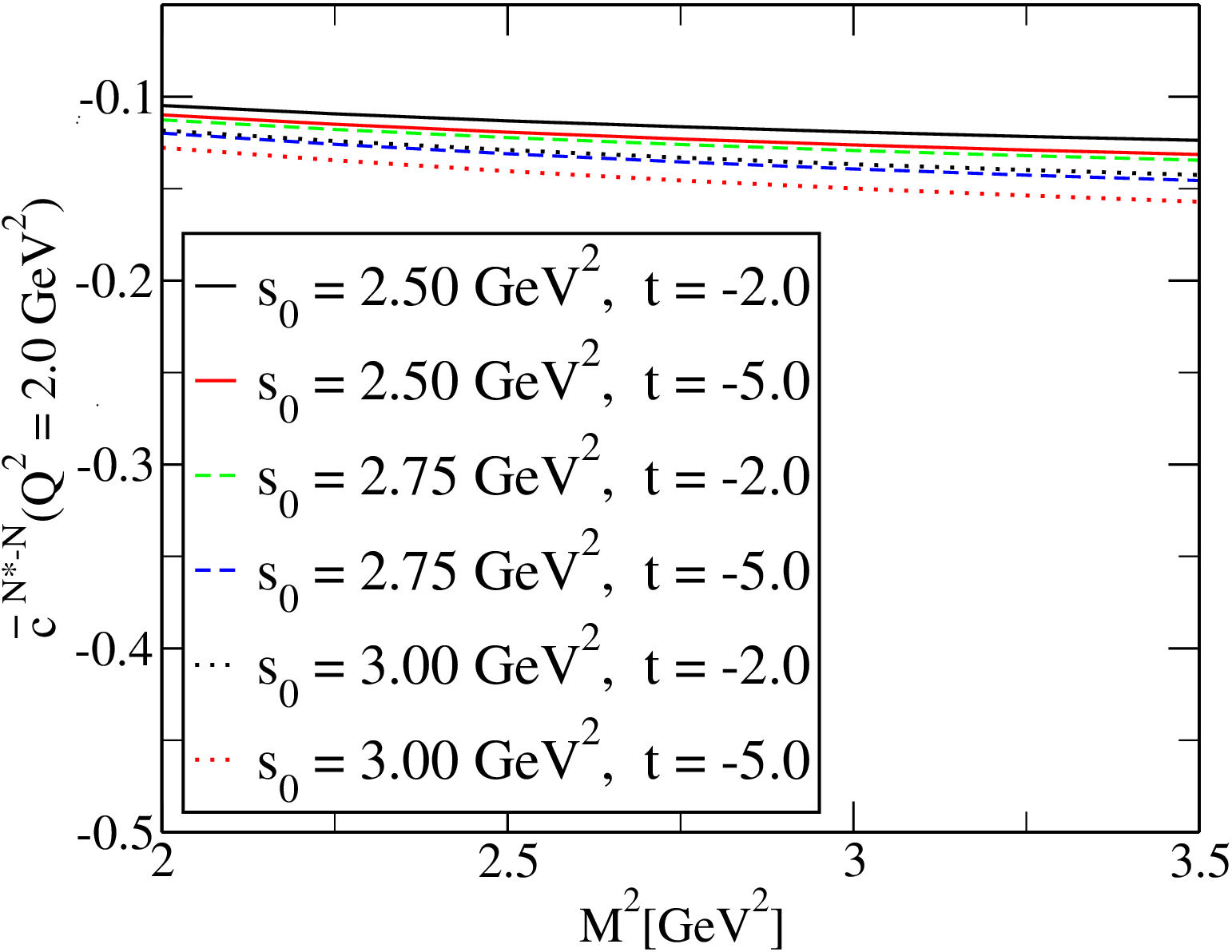}}
 \caption{The dependence of the the GFFs of $N^*\rightarrow N$ transition on Borel mass parameter $M^2$ at $Q^2$ = 2.0~GeV$^2$ and various values of continuum threshold $ s_0 $ and arbitrary mixing parameter $ t $ at their working region:
 (a), (c), (e) and (g) for the LCSR-1;  (b), (d) and  (f) for the LCSR-2. }
 \label{Msqfigs}
  \end{figure} 
  \begin{figure}[htp]
\centering
 \subfloat[]{\includegraphics[width=0.3\textwidth]{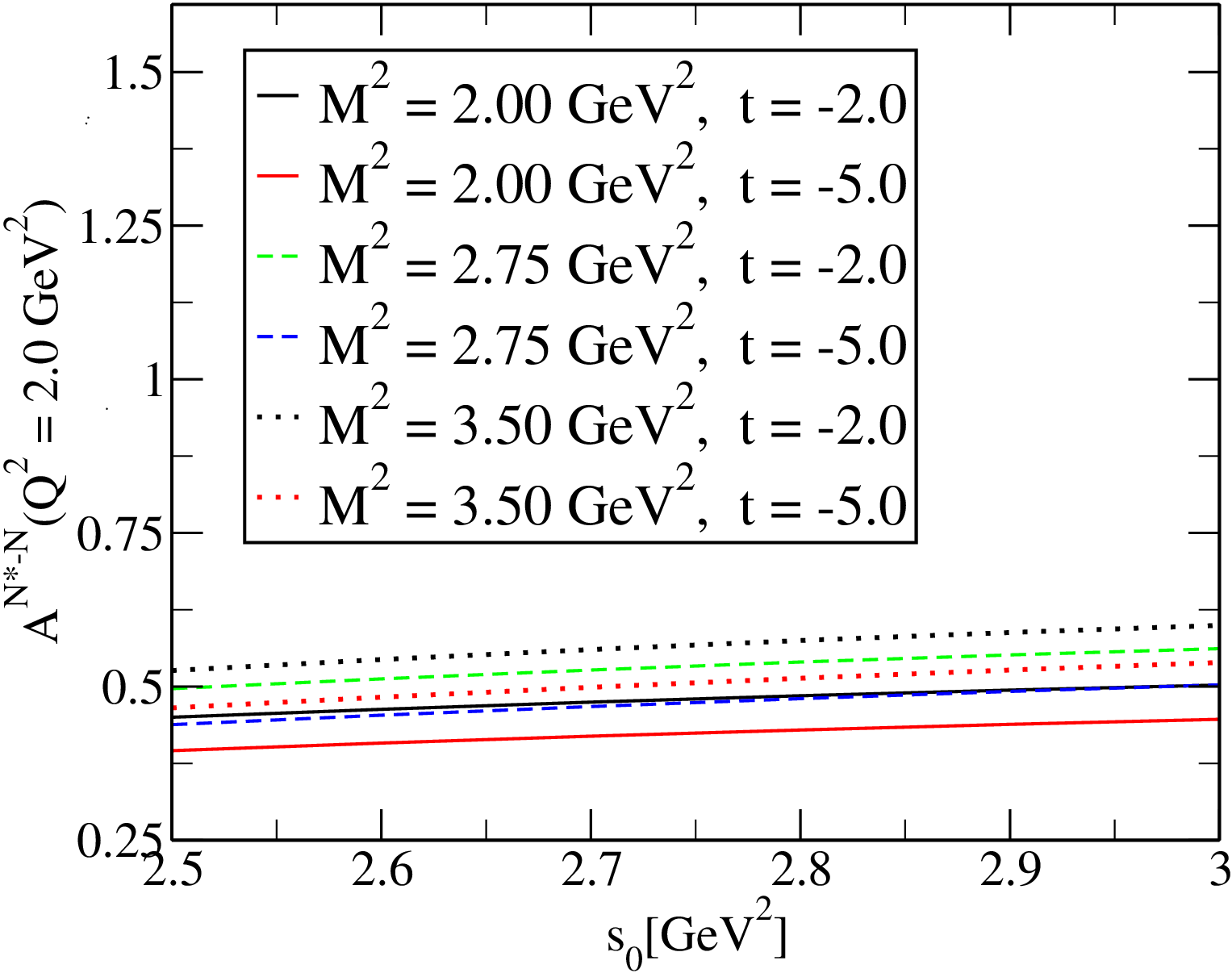}}~~~~~~~~~
 \subfloat[]{\includegraphics[width=0.3\textwidth]{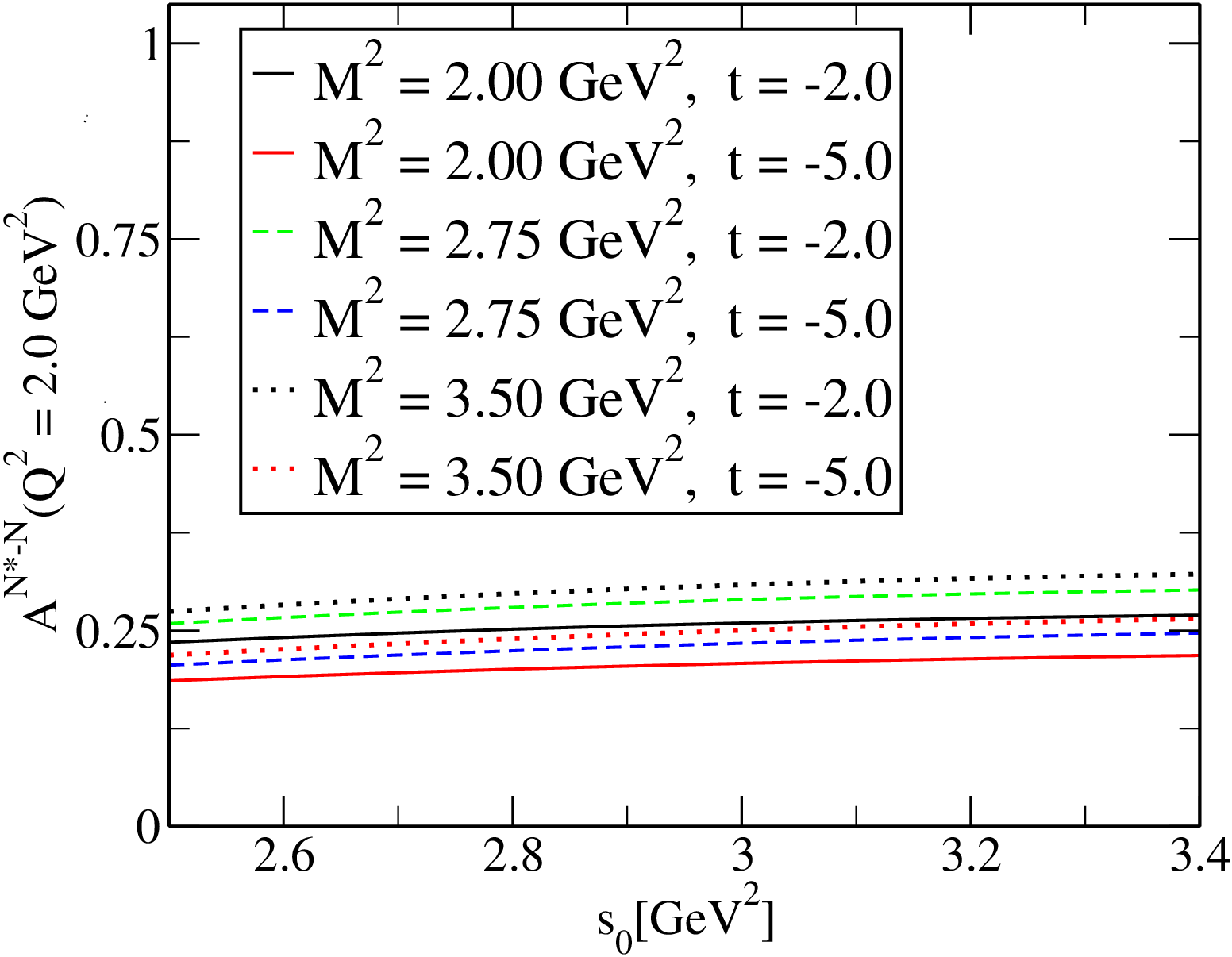}}\\
    \vspace{0.2cm}
    \subfloat[]{\includegraphics[width=0.3\textwidth]{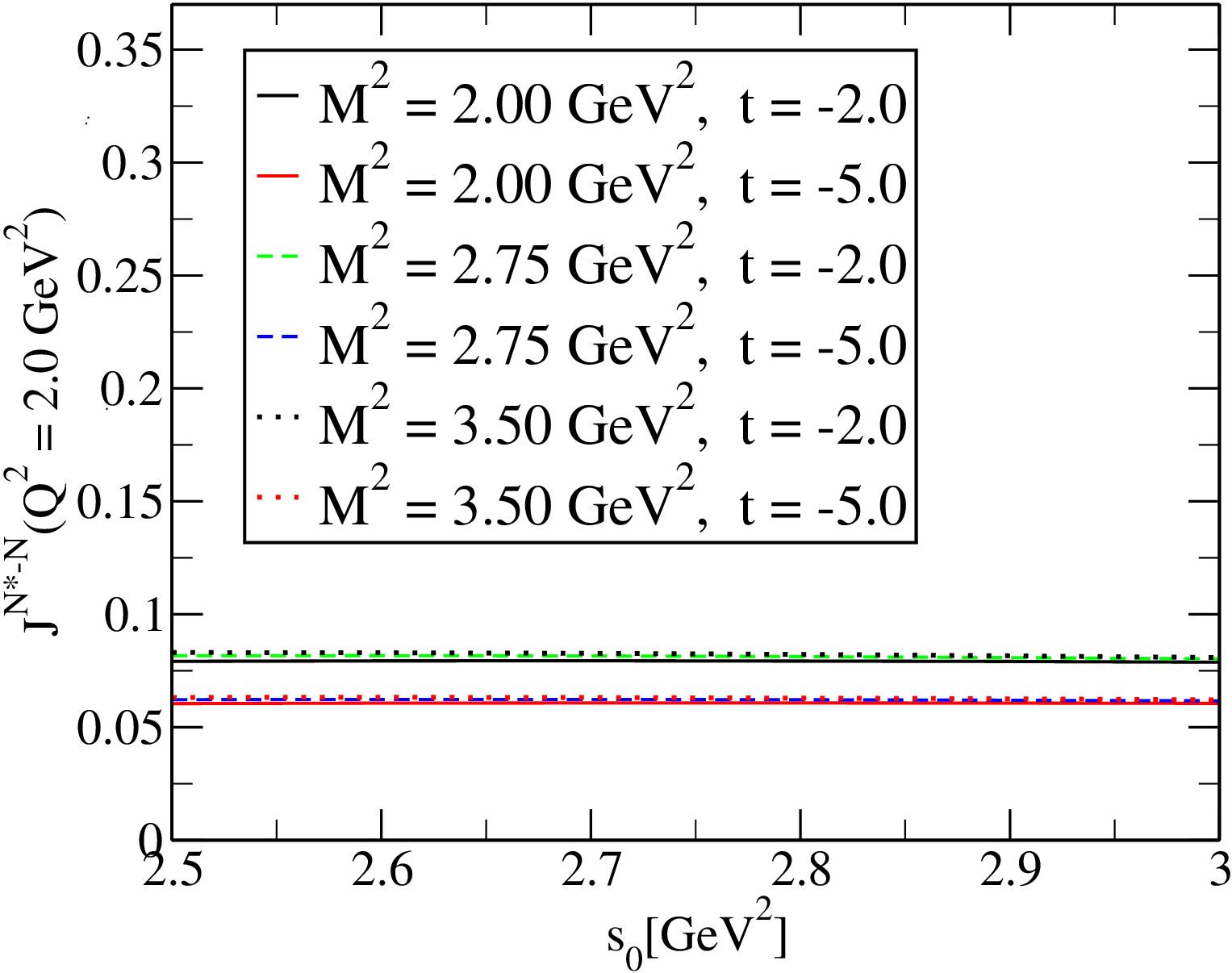}}~~~~~~~~~
    \subfloat[]{\includegraphics[width=0.3\textwidth]{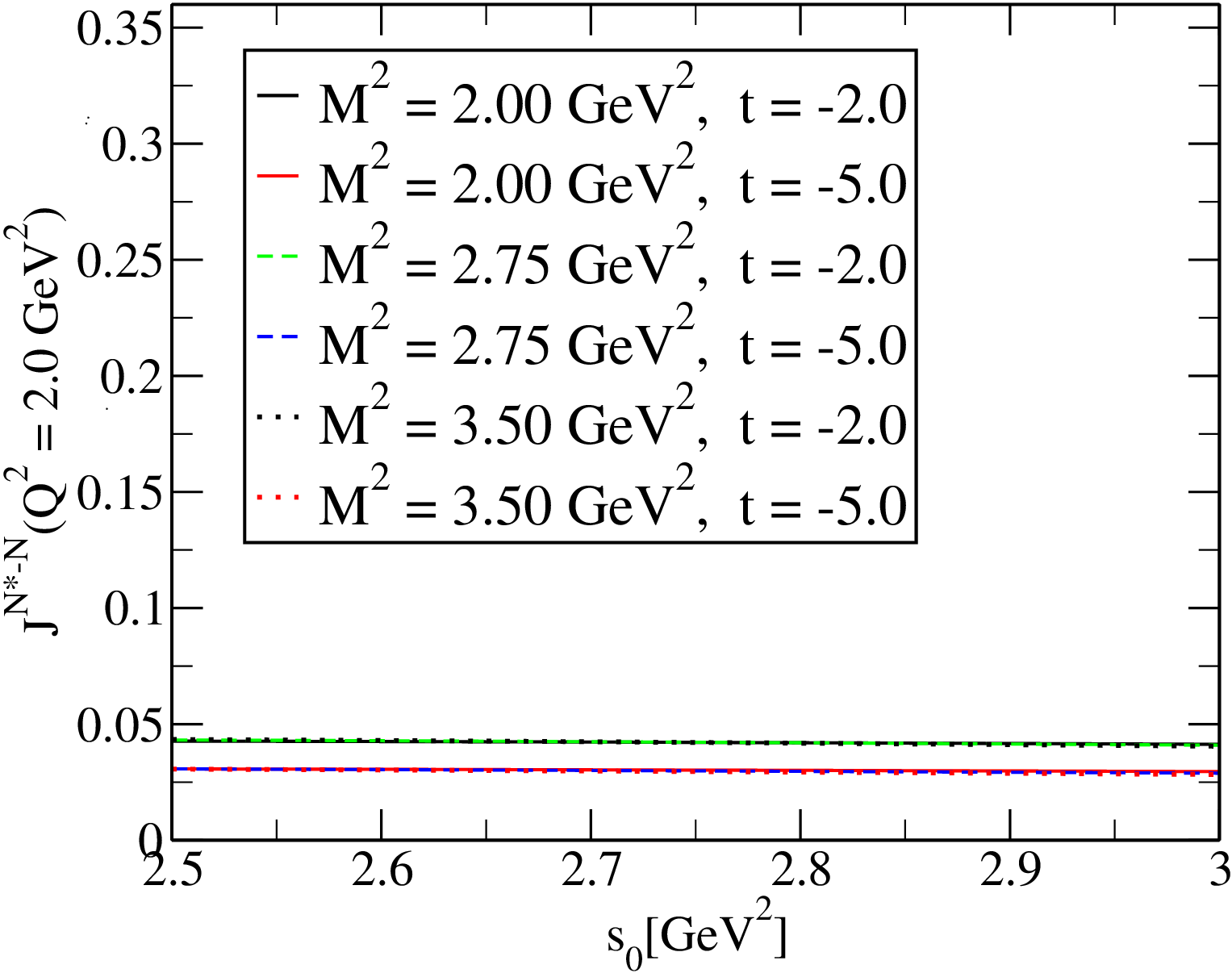}}\\
 \vspace{0.2cm}
    \subfloat[]{\includegraphics[width=0.3\textwidth]{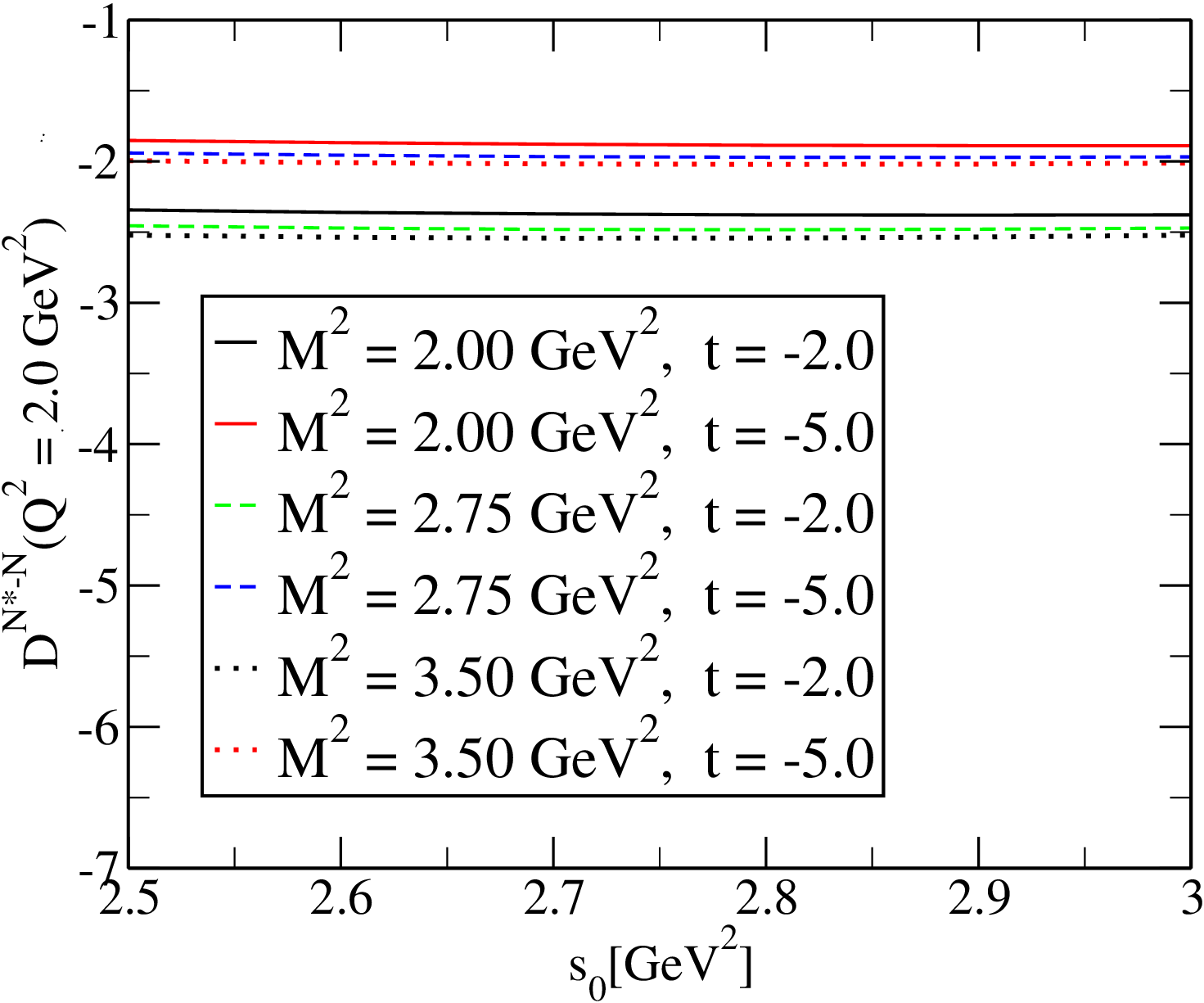}}~~~~~~~~~
    \subfloat[]{\includegraphics[width=0.3\textwidth]{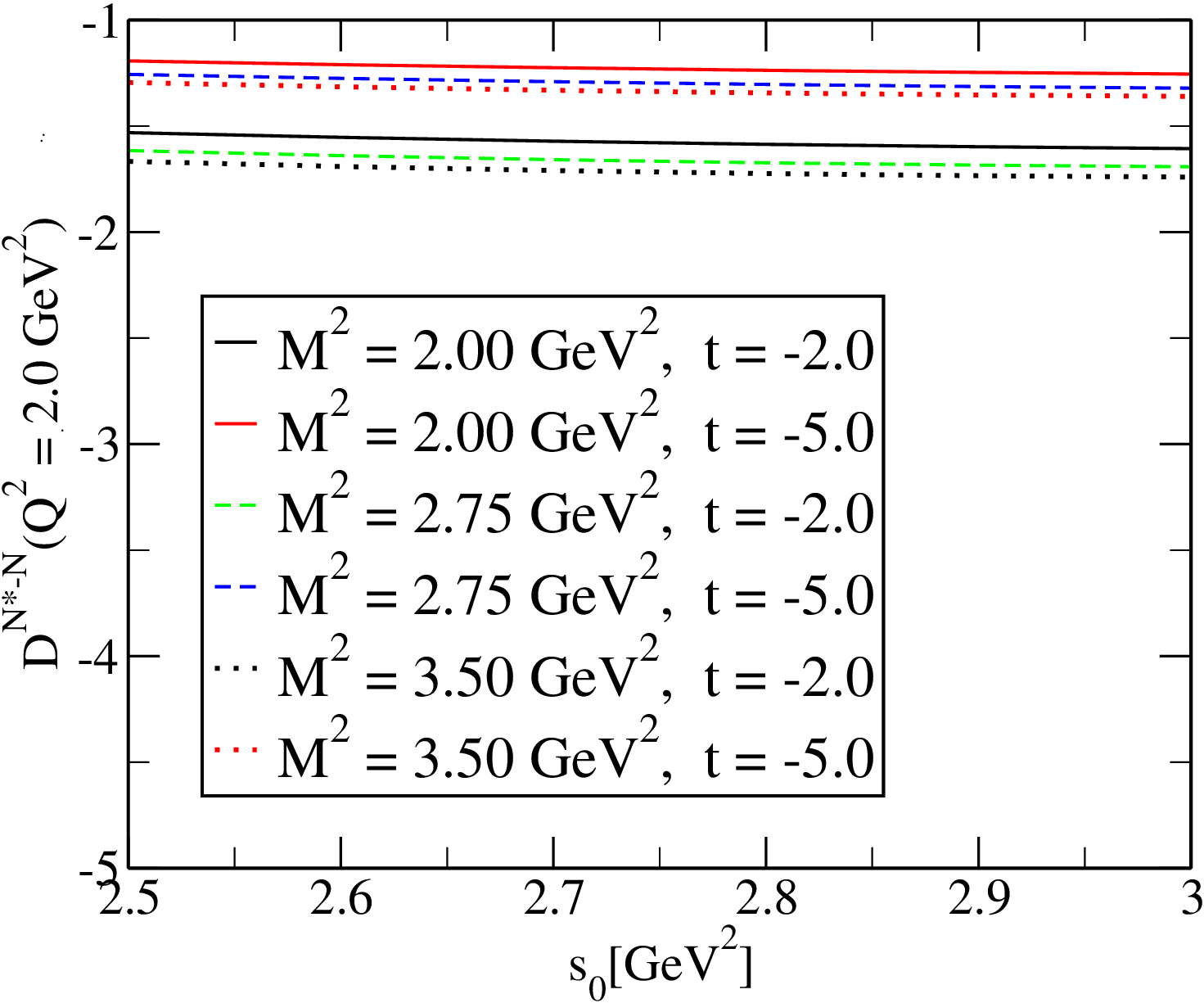}}\\
   \vspace{0.2cm}
 \subfloat[]{\includegraphics[width=0.3\textwidth]{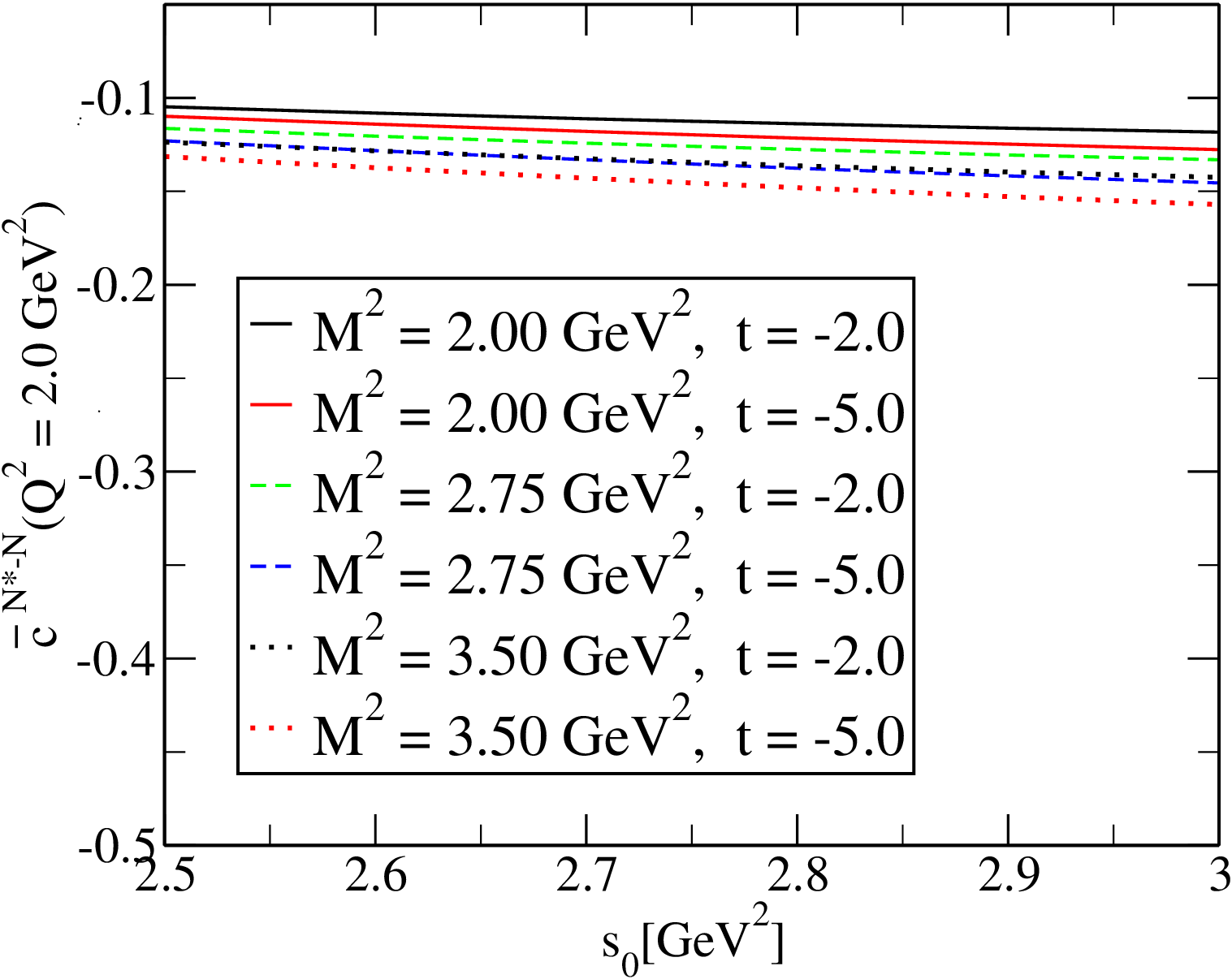}}
 \caption{The dependence of the the GFFs of $N^* \rightarrow N$ transition on continuum threshold $s_0$ at $Q^2$ = 2.0~GeV$^2$ and various values of Borel mass parameter $ M^2 $ and arbitrary mixing parameter $ t $ at their working region: 
(a), (c), (e) and (g) for the LCSR-1; 
 (b), (d) and  (f)  for the LCSR-2.}
 \label{s0sqfigs}
  \end{figure}
\begin{figure}[htp]
\centering
 \subfloat[]{\includegraphics[width=0.3\textwidth]{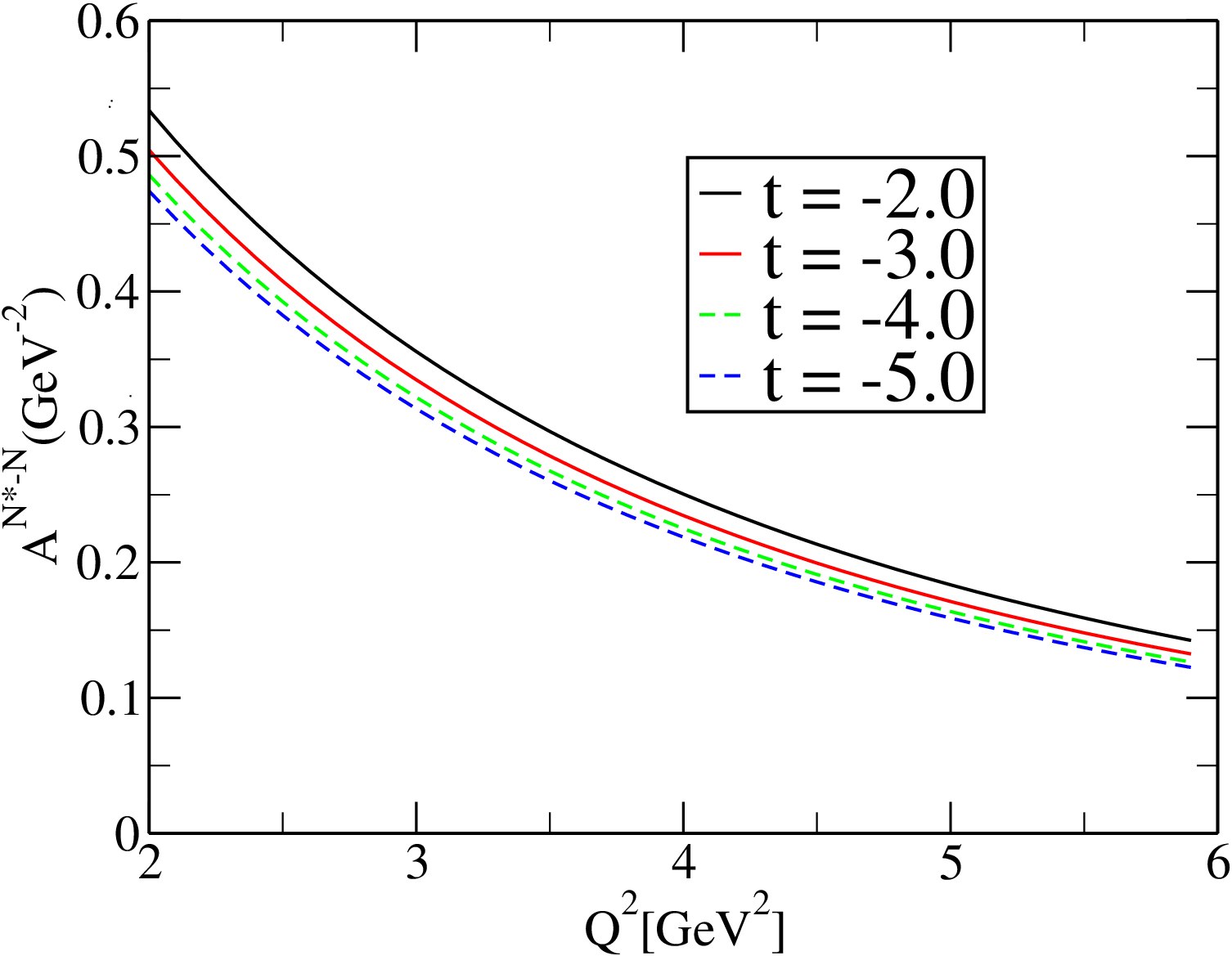}}~~~~~~~~~
 \subfloat[]{\includegraphics[width=0.3\textwidth]{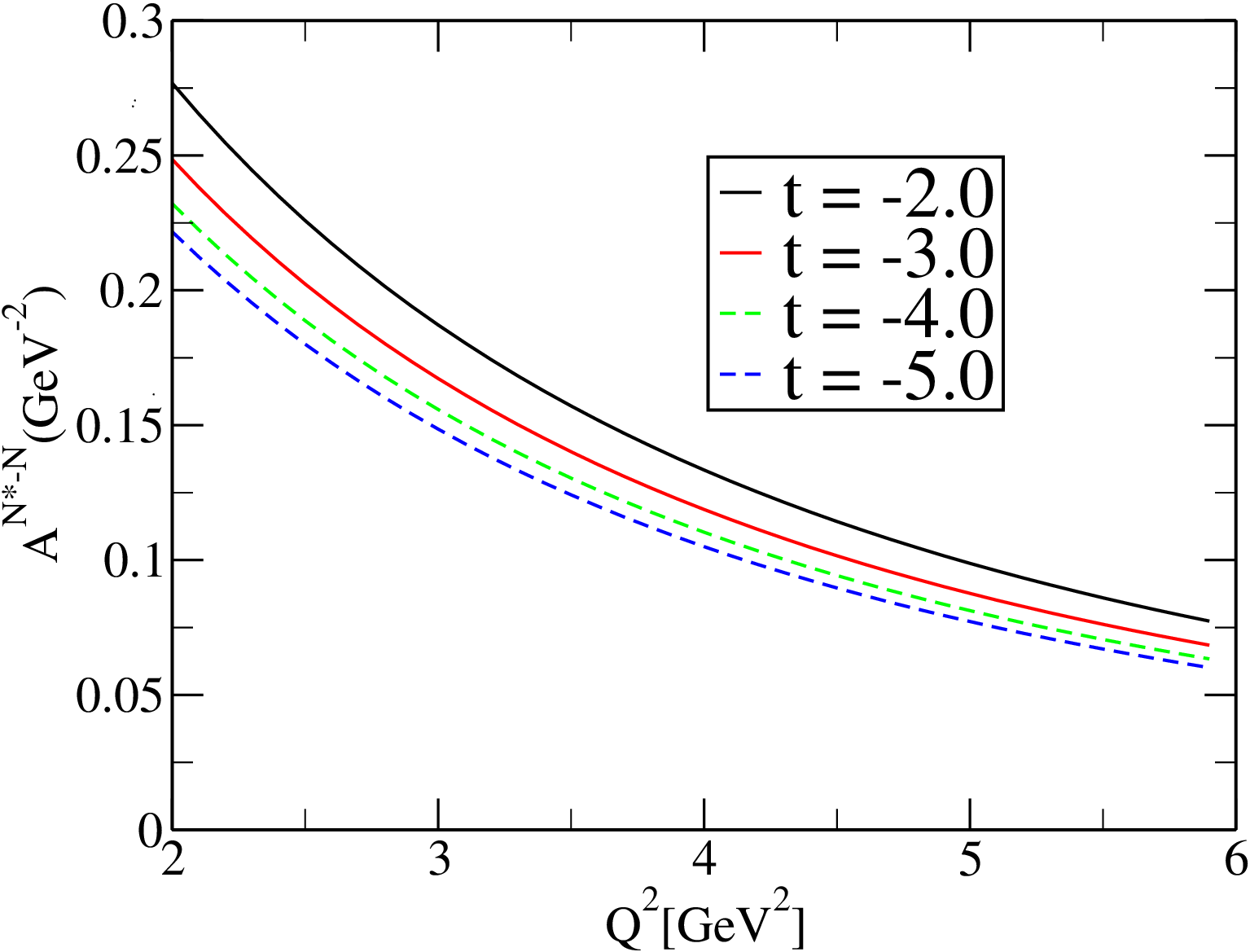}}\\
    \vspace{0.2cm}
    \subfloat[]{\includegraphics[width=0.3\textwidth]{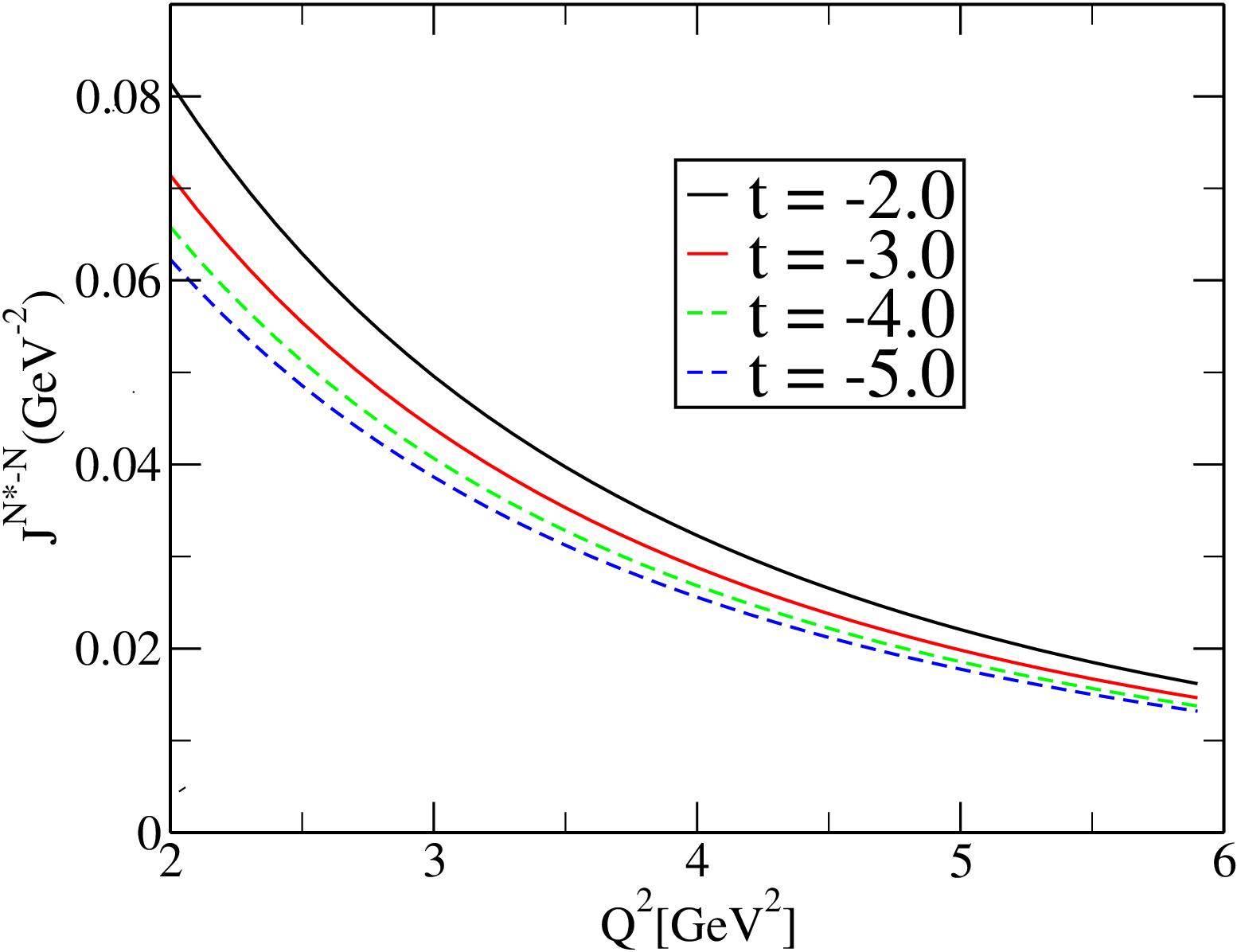}}~~~~~~~~~
    \subfloat[]{\includegraphics[width=0.3\textwidth]{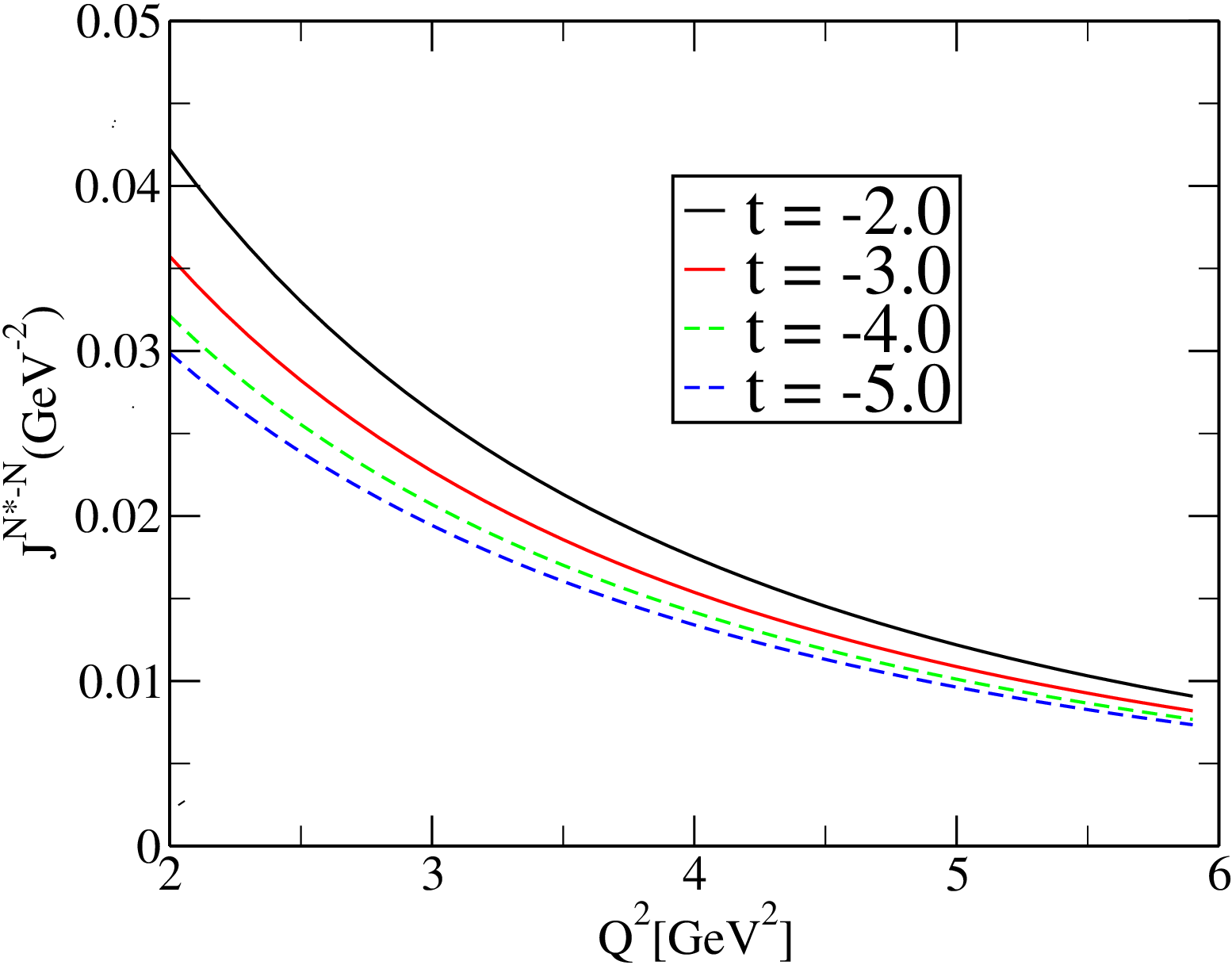}}\\
 \vspace{0.2cm}
    \subfloat[]{\includegraphics[width=0.3\textwidth]{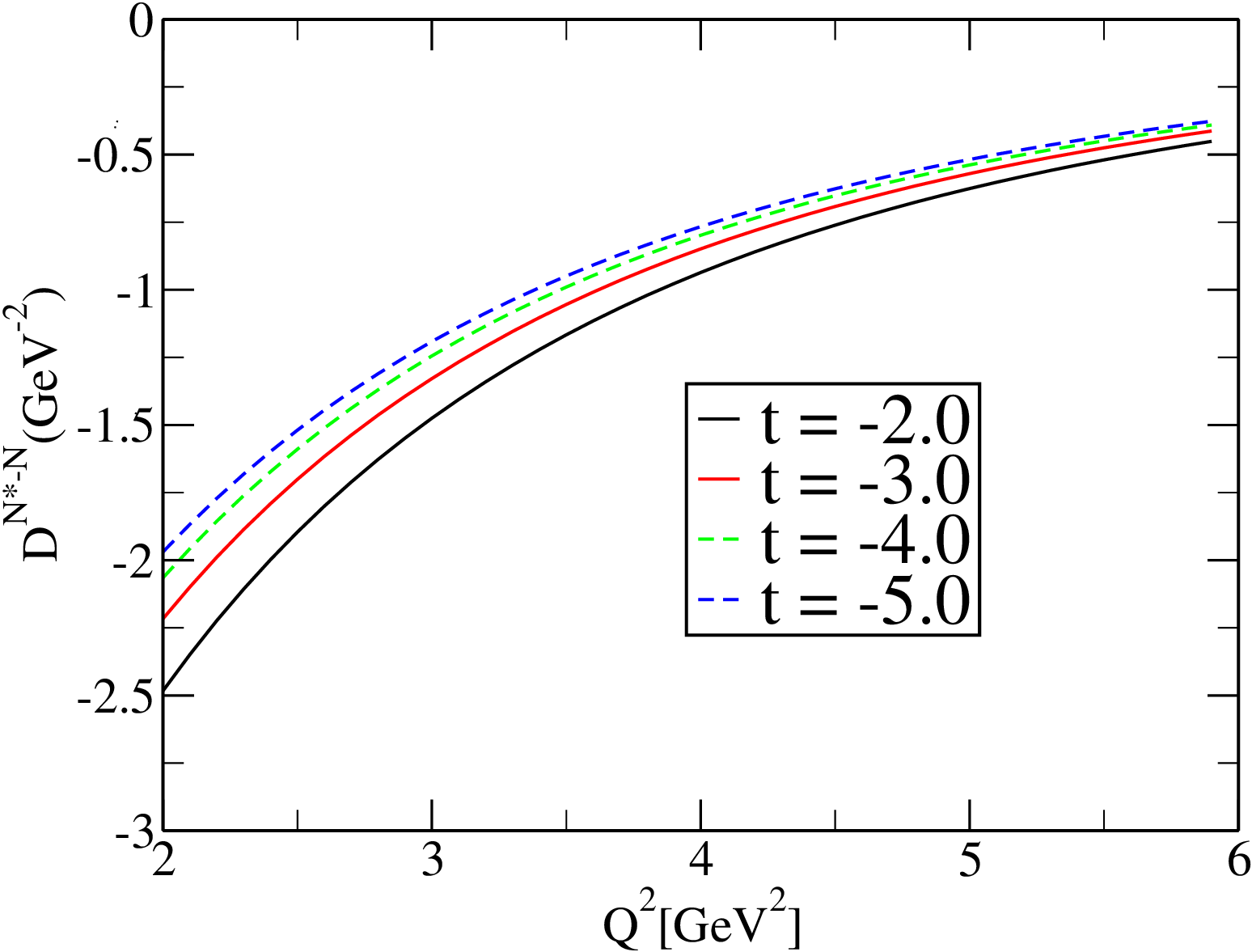}}~~~~~~~~~
     \subfloat[]{\includegraphics[width=0.3\textwidth]{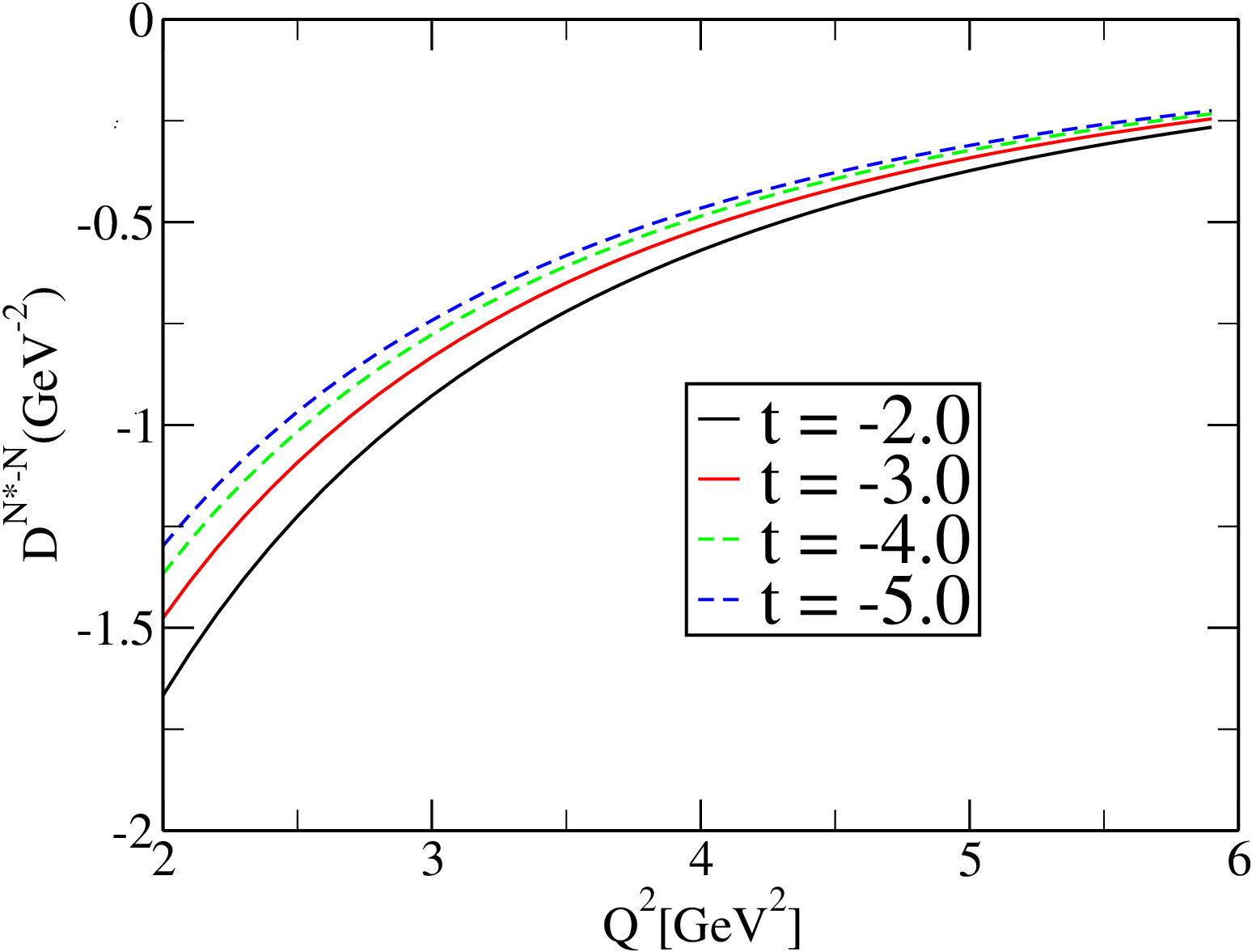}}\\
   \vspace{0.2cm}
 \subfloat[]{\includegraphics[width=0.3\textwidth]{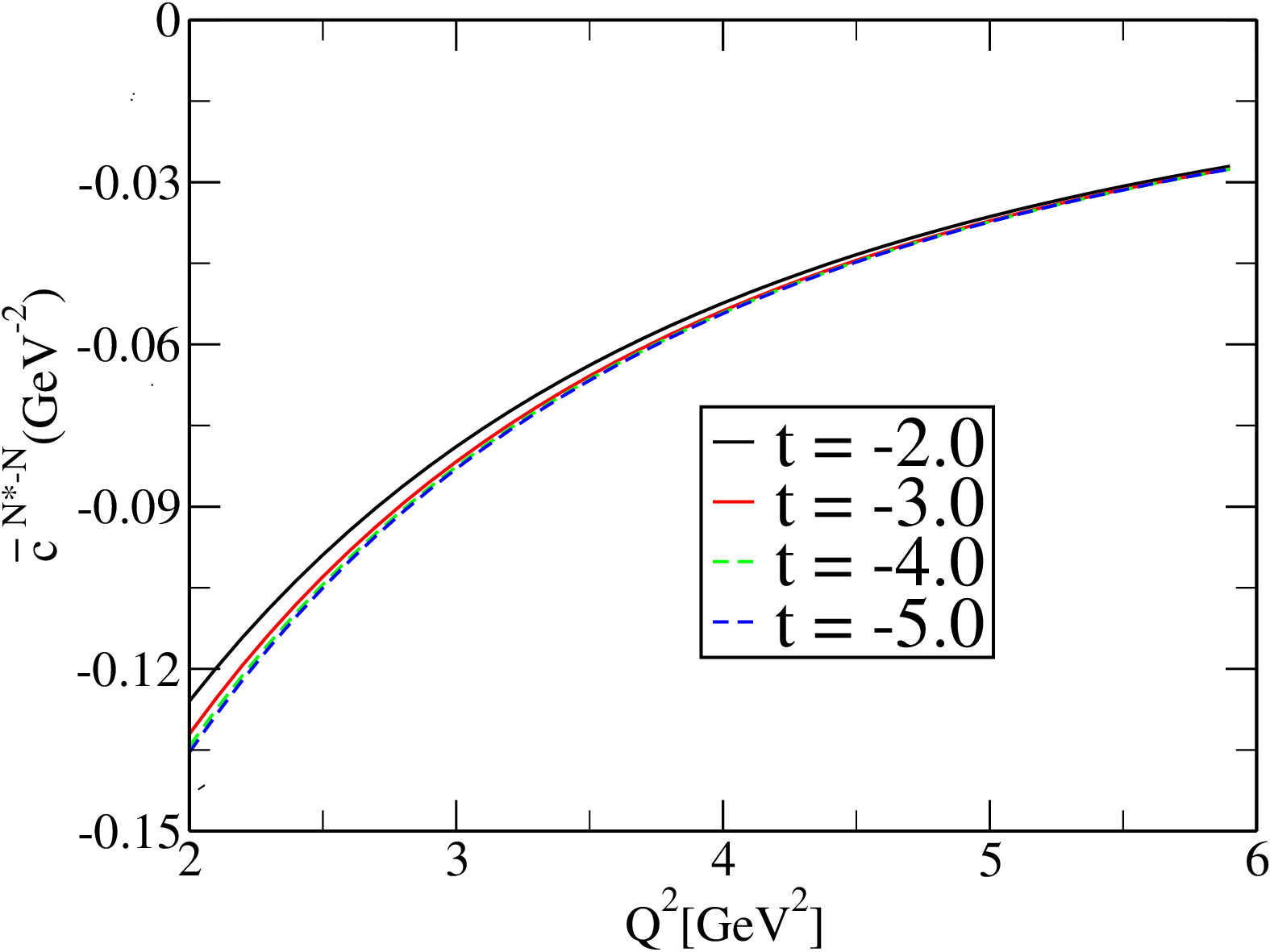}}
 \caption{The dependence of the GFFs of $N^*\rightarrow N$ transition on $Q^2$ at $M^2$ = 2.75~GeV$^2$, $s_0$ = 2.75~GeV$^2$ and 
  various values of arbitrary mixing parameter $ t $: 
(a), (c), (e) and (g) for the LCSR-1; 
 (b), (d) and  (f)  for the LCSR-2.}
 \label{Qsqfigs}
  \end{figure}
  
\end{widetext}

\newpage
\bibliography{refs}

\end{document}